\documentclass[amsbook,amsmath,amssymb,byrevtex,dvips,showpacs]{revtex4}
\usepackage{graphicx,amssymb,latexsym,amsmath,amsbsy}

\begin{document}

\title{Quantum Monte Carlo Simulations of the BCS-BEC Crossover at Finite Temperature}
\author{ Aurel Bulgac$^1$, Joaqu\'{\i}n E. Drut$^1$\footnote{Address as of April 2008: {\it Department of Physics, The Ohio State University, Columbus, OH 43210, USA.}} and Piotr Magierski$^2$}
\affiliation{$^1$Department of Physics, University of
Washington, Seattle, WA 98195--1560, USA}
\affiliation{$^2$Faculty of Physics, Warsaw University of Technology,
ulica Koszykowa 75, 00-662 Warsaw, POLAND }

\begin{abstract}
The Quantum Monte Carlo method for spin 1/2 fermions at finite temperature is formulated for dilute systems with an s-wave interaction. The motivation and the formalism are discussed along with descriptions of the algorithm and various numerical issues. We report on results for the energy, entropy and chemical potential as a function of temperature. We give upper bounds on the critical temperature $T_c$ for the onset of superfluidity, obtained by studying the finite size scaling of the condensate fraction. All of these quantities were computed for couplings around the unitary regime in the range $-0.5 \le (k_F a)^{-1} \le 0.2$, where $a$ is the s-wave scattering length and $k_F$ is the Fermi momentum of a non-interacting gas at the same density. In all cases our data is consistent with normal Fermi gas behavior above a characteristic temperature $T_0 > T_c$, which depends on the coupling and is obtained by studying the deviation of the caloric curve from that of a free Fermi gas. For $T_c < T < T_0$ we find deviations from normal Fermi gas behavior that can be attributed to pairing effects. Low temperature results for the energy and the pairing gap are shown and compared with Green Function Monte Carlo results by other groups.
\end{abstract}

\date{\today}

\pacs{03.75.Ss, 03.75.Hh, 05.30.Fk}


\maketitle

\numberwithin{equation}{section} 

\section{Introduction}
The last few years have witnessed an extraordinary progress in the field of cold
fermionic atoms, particularly since the experimental observation of superfluidity in these 
systems \cite{exp}. Ultracold atomic gases provide an exceptional opportunity to investigate strongly correlated fermions. Taking advantage of Feshbach resonances, experimentalists vary the strength of the interaction between the atoms at will, offering an unprecedented ability to study the BCS-BEC crossover.  
It is by now well established that, for the case of broad resonances, the physics of these systems is chiefly captured by a model of spin 1/2 fermions with a contact s-wave attractive interaction. On the theoretical side, our overall understanding of these remarkable many-body systems has improved dramatically, although many questions still remain unanswered. Systems in the BCS-BEC crossover are strongly correlated, and nonperturbative approaches are needed. As a consequence, numerical simulations of spin 1/2 fermions at zero and non-zero temperature have lately attracted extraordinary attention (see refs.\cite{carlson, chang, giorgini, stefano, bdm, burovski, njp, lee}). 

Of particular interest within the BCS-BEC crossover is the so-called unitary regime (see \cite{gfb, baker, ho, CarlsonReddy}). This regime is formally defined as the limit of diluteness with respect to the range of the interaction $r_0$, and large scattering length $a$, such that $nr_0^3 \ll 1 \ll n|a|^3$, where $n$ is the particle number density. The thermal behavior in the crossover is characterized by a dimensionless universal function conventionally called $\xi(T/\varepsilon_F,1/k_F a)$, that depends on the temperature $T$ (in units of the free gas Fermi energy $\varepsilon_F = \hbar^2 k_F^2/2m$, where $k_F = (3\pi^2n)^{1/3}$), and the strength of the interaction, usually parameterized by $(k_Fa)^{-1}$. Throughout this work we shall use units in which Boltzmann's constant is $k_B = 1$. The function $\xi$ represents the ratio of the energy $E$ to the energy of a free Fermi gas at the same density $E_F = 3/5 N \varepsilon_F$. The value of $\xi$ at unitarity and at $T = 0$, which we shall denote $\xi_s$, has been determined approximately by various authors, and recent Quantum Monte Carlo calculations and extrapolations to zero range yield $\xi_s = 0.40(1)$ (see \cite{GezerlisCarlson} and references therein). Besides ultracold atomic gases, the unitary regime is relevant for dilute neutron matter, although in that case finite effective range effects cannot be neglected \cite{schwenk}.

In this paper the determination of the universal function $\xi(T/\varepsilon_F, 1/k_F a)$, along with other thermodynamic quantities (including the critical temperature $T_c$ for the onset of superfluidity), will constitute our main results. We explore the unitary limit, where $k_F |a| \rightarrow \infty$, as well as the case of finite scattering length in the range $-0.5 \le (k_F a)^{-1} \le 0.2$. The paper is organized as follows: in section II we formulate the problem and describe the nonperturbative
technique based on the discretization of the space-time and subsequent evaluation of
the thermodynamic quantities through a Quantum Monte Carlo simulation. In section III we describe the numerical and computational techniques that were used. The main results are discussed in sections IV (at unitarity) and V (away from unitarity), and the final conclusions in section VI.

\section{Mathematical formulation: Fermions on a space-time lattice}

\subsection{The Hamiltonian and the running coupling constant}

\begin{figure}[htb]
\centering
\includegraphics[width=9cm]{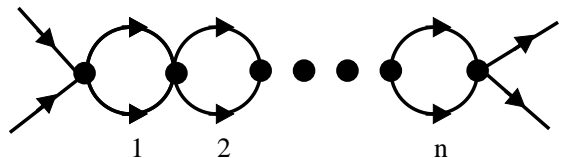}
\caption{\label{bubblenth} n-th term in the bubble sum.}
\end{figure}

The interaction that captures the physics of a dilute, unpolarized system of fermions
is a zero-range two-body interaction 
$V(\mathbf{r}_1 - \mathbf{r}_2) = -g \delta(\mathbf{r}_1 - \mathbf{r}_2)$.
The Hamiltonian of the
system in the second quantization representation reads:
\begin{equation}
\hat{H}= 
\int d^{3}r \left (
 -\sum_{\sigma=\uparrow,\downarrow}
  \hat{\psi}_{\sigma}^{+}(\mathbf{r})\frac{\hbar^2\nabla^2}{2m}\hat{\psi}_{\sigma} (\mathbf{r})
 - g \hat{n}_\uparrow (\mathbf{r}) \hat{n}_\downarrow (\mathbf{r}) \right ),
\end{equation}
where $\hat{n}_\sigma (\mathbf{r})=\hat{\psi}_{\sigma}^{+}(\mathbf{r})\hat{\psi}_{\sigma} (\mathbf{r})$.
This interaction requires that we specify a regularization procedure, which we do by introducing a momentum cut-off $\hbar k_c$ (thus requiring all two-body matrix elements to vanish if the relative momentum of the incoming particles exceeds the cut-off). Once this cut-off is imposed, the value of the bare coupling $g$ can be tuned to fix the value of the physical renormalized coupling, which in this case will be the s-wave scattering length $a$. Indeed, the diagonal $T$ matrix describing two-particle scattering induced by the interaction takes the simple form:
\begin{equation}
T(k)=\frac{g}{(2\pi)^3}
\sum_{n=0}^{\infty}\left [g \int\frac{d^3 k'}{(2\pi)^3}
\left({\frac{\hbar^2 k^{2}}{m} - \frac{\hbar^2 k'^{2}}{m}+ i0^{+}}\right)^{-1}\right ]^{n} 
= -\frac{g}{(2\pi)^3}
\left[ 1+\frac{g m}{2\pi^2\hbar^2}\left (-k_{c} + k~\mbox{atanh} \left (\frac{k_{c}}{k} \right) \right ) \right]^{-1},
\end{equation}
which is equivalent to finding the vacuum 4-point amplitude and determining the scattering length by summing all the `bubble' diagrams (see Fig.~\ref{bubblenth}). 
The low-momentum expansion of the scattering amplitude reads:
\begin{equation}
f(k)\approx \left[{-ik+\frac{4\pi\hbar^{2}}{gm}-\frac{2k_{c}}{\pi}+\frac{2k^{2}}{\pi k_{c}} + O(k^{3})}\right]^{-1}.
\end{equation}
At low momentum we have, by definition, $f(k) = [-ik - 1/{a} + r_\text{eff}k^2/2 + O(k^3)]^{-1}$, which gives the relation between the bare coupling constant $g$ and the scattering length $a$ at a given momentum cutoff $\hbar k_c$:
\begin{equation}
\label{RenormalizedCoupling}
\frac{1}{g} = -\frac{m}{4 \pi \hbar^2 a} + \frac{k_c m}{2 \pi^2 \hbar^2}.
\end{equation} 
Note that an effective range $r_\text{eff}$ is generated which is independent of the coupling constant
$r_\text{eff}=\displaystyle{\frac{4}{\pi k_{c}}}$.

\subsection{Discrete variable representation}

To determine the thermal properties of spin 1/2 fermions in a non-perturbative manner, 
we have placed the system on a three dimensional (3D) cubic spatial lattice with periodic boundary 
conditions. The lattice spacing $l$ and size $L = N_s l$ introduce natural ultraviolet (UV) and 
infrared (IR) momentum cut-offs given by $\hbar k_c=\pi\hbar/l$ and $\hbar\Lambda_0=2\pi\hbar/L$, respectively. 
The momentum space has the shape of a cubic lattice, 
with size $2\pi\hbar/l$ and spacing $2\pi\hbar/L$.
To simplify the analysis, however, we place a spherically symmetric UV cut-off, including  
only momenta satisfying $k \le k_c \le \pi/l$.

The discretization transforms the functions of continuous variables into functions 
defined on a discrete set of coordinate values:
\begin{eqnarray}
\hat{\psi}_{\sigma}(\mathbf{r}_{i}) \rightarrow \hat{\psi}_{\sigma}({\bf i}) \\
\hat{\psi}_{\sigma}^{+}(\mathbf{r}_{i}) \rightarrow \hat{\psi}_{\sigma}^{+}({\bf i}) \\
\hat{n}_{\sigma}(\mathbf{r}_{i}) \rightarrow \hat{n}_{\sigma}({\bf i}),
\end{eqnarray}
where $\mathbf{r}_{i}={\bf i} l$ and ${\bf i}=(i_x ,i_y ,i_z )$ denotes the lattice sites,
and $i_{x},i_{y},i_{z}=1,...,N_s$.
The discretization can affect the accuracy of the obtained results and therefore requires
careful analysis. We address this issue by discussing the so-called discrete variable representation 
(DVR) basis sets, which is the underlying framework of our lattice approach, see Ref.~\cite{little}.

Let us call ${\cal H} = L^2(M)$ the Hilbert space of our problem, i.e. the set of square-integrable wave functions on the manifold $M$. Let $P$ be a projector defined on $\cal H$
such that ${\cal S} = P \cal H$ is the projected subspace. Given a set of $N$ grid points $\{x_\alpha, \alpha = 0,...,N-1\}$ in $d$ dimensions one can define projected $\delta$-functions:
$\Delta_\alpha(x) = P[\delta(x-x_\alpha)]$. Alternatively $|\Delta_\alpha \rangle = P |x_\alpha \rangle$ or
$\Delta_\alpha(x)=\langle x|\Delta_\alpha\rangle$, in Dirac notation.

It follows that 
\begin{equation}
\langle \Delta_\alpha|\Delta_\beta\rangle = \Delta_\beta(x_\alpha) = \Delta^*_\alpha(x_\beta) 
\end{equation}
and as a result the set of projected $\delta$-functions $\{\Delta_\alpha, \alpha = 0,...,N-1\}$ is
orthogonal if and only if 
\begin{equation}
\Delta_\alpha(x_\beta) = K_\alpha \delta_{\alpha \beta},
\end{equation}
where $K_\alpha = \langle \Delta_\alpha|\Delta_\alpha\rangle$, and we can normalize
the projected functions to get their orthonormalized version:
\begin{equation}
| F_\alpha \rangle = \frac{1}{\sqrt{K_\alpha}}| \Delta_\alpha\rangle.
\end{equation}
Given a wave-function $\psi \in \cal S$, an expansion of the form
\begin{equation}
\psi(x) = \sum_{\alpha} c_\alpha F_\alpha(x)
\end{equation}
exists, and the coefficients are given by the values of $\psi(x)$ at the lattice sites $x_\alpha$:
\begin{equation}
c_\alpha = \int dxF_\alpha^*(x) \psi(x) = \frac{1}{\sqrt{K_\alpha}}\psi(x_\alpha)
\end{equation}

\begin{figure}[htb]
\includegraphics[scale=0.4]{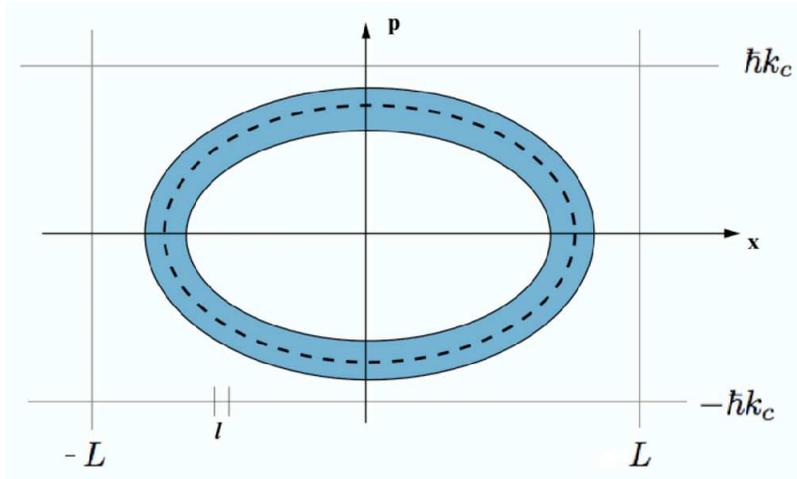}
\caption{\label{phasespacediagram} (Color online) Representation of phase space. The dashed line shows a classical trajectory, while the shaded region represents the quantum fluctuations around such trajectory. The horizontal and vertical lines show the cutoffs in real space $L$, and in momentum space $\hbar k_c = \hbar \pi / l$, where $l$ is the lattice spacing (assuming the lattice is a uniform square lattice). Functions within the larger oval region are well described by a given basis if the cutoffs enclose that region, as shown in this figure.}
\end{figure}

On the other hand, if $\psi(x)$ is not fully contained in $\cal S$, then our basis set
will not be sufficiently rich to represent it. However, if the semiclassical region of phase space that we wish to represent (see Fig.~\ref{phasespacediagram}) is contained in $\cal S$ (and in particular if the UV momentum cutoff $\hbar k_c $ is larger than the highest momentum we wish to represent, or equivalently if the lattice spacing $l$ is chosen to be sufficiently small), then it can be shown (see Ref. \cite{little}) that the errors are exponentially suppressed. 

The representation of the Hamiltonian on the lattice has been obtained
by noticing that two terms representing the kinetic and interaction energy
are local in momentum and coordinate spaces, respectively. 
Thus, the kinetic term reads:
\begin{equation}
\hat{K}= l^{-3}\sum_{\sigma=\uparrow,\downarrow}\sum_{\bf \tilde{i}}\frac{\hbar^2 k_{\tilde{i}}^2}{2m}
        \hat{n}_{\sigma} ({\bf \tilde{i}}),
\end{equation}
where ${\bf \tilde{i}}=(\tilde{i}_x,\tilde{i}_y,\tilde{i}_z)$ enumerates the lattice
sites in the momentum space, $\tilde{i}_x,\tilde{i}_y,\tilde{i}_z=0,\pm 1,\pm 2,...,N_s/2 (\mbox{or} -N_s/2)$ and ${\bf k_{\tilde{i}}}={\bf \tilde{i}}{2\pi}/{L}$.
The operator $\hat{n}_{\sigma} ({\bf \tilde{i}})$ denotes the occupation number operator of the single particle state with momentum $\hbar {\bf k_{\tilde{i}}}$ and spin $\sigma$. On the other hand, the interaction becomes a simple Hubbard attractive potential 
\begin{equation}
\hat{V} = -g l^{3} \sum_{\bf i} \hat{n}_\uparrow ({\bf i}) \hat{n}_\downarrow ({\bf i}).
\end{equation}
From this point on we shall omit any factors of $l^{3}$ or $l^{-3}$ coming from the volume elements in real and momentum space, which amounts to a specific choice of units. If the cutoff is chosen to be $k_c = {\pi}/{l}$, then infinite scattering length corresponds to a coupling given by $g = 2 \pi \hbar^2 l /m$ (see Eq.~(\ref{RenormalizedCoupling})).

In order to avoid numerical inaccuracies associated with the discretization
of the differential operators on the lattice, we use both momentum
and coordinate representation of the lattice and a Fast Fourier Transform (FFT) 
to switch between the two. In the following we shall rename the indices ${\bf \tilde{i}}$ and {\bf i} in favor of the more suggestive names ${\bf k}$ and ${\bf r}$, respectively.

\subsection{Discrete auxiliary fields and positivity of the probability measure}

In order to study the thermal properties we chose the grand canonical ensemble, where the thermodynamic variables are the temperature $T$, the chemical potential $\mu$ and the volume $V$. The partition function and the average of an observable $\hat{O}$ are calculated according to

\begin{eqnarray}
Z(\beta,\mu,V) &=&  {\mathrm{Tr}} \left \{ \exp [-\beta (\hat{H}-\mu \hat{N})] \right \} ,
\nonumber \\
O(\beta,\mu,V) &=& 
\frac{{\mathrm{Tr}} \; \left \{ \hat{O}\exp [-\beta (\hat{H}-\mu \hat{N})] \right \}}{Z(\beta,\mu,V)} ,
\end{eqnarray}
where $\beta = 1/T$ (as mentioned before, in this work we will take Boltzmann's constant to be $k_B = 1$).
%
By factorizing the statistical weight using the Trotter formula, one obtains
\begin{equation}
\exp [-\beta (\hat{H}-\mu \hat{N})] = \prod_{j=1}^{N_\tau}\exp [-\tau (\hat{H}-\mu \hat{N})]
\end{equation}
where $\beta = N_\tau \tau$. The next step is to decompose the exponentials on the right hand side into exponentials that depend on the kinetic and potential energy operators separately. This can be achieved to a suitable order through the factorization
\begin{equation}
\exp [-\tau (\hat{A}+\hat{B})]\simeq\prod_{k=1}^{M}\exp [- a_{k} \tau \hat{A}] 
                                              \exp [- b_{k} \tau \hat{B}],
\end{equation}
where the coefficients $a_{k}, b_{k}$ are determined by the required order of accuracy
\cite{decomp}. However in order to obtain stable numerical results all of the coefficients
have to be positive. This poses a practical limitation on the above formula which works
only up to the second order, for which the expansion is
\begin{eqnarray}\label{eq:fact-Exp}
\exp[ -\tau(\hat{H}-\mu \hat{N})] =
\exp \left [ -\frac{\tau (\hat{K}-\mu \hat{N})}{2}\right ]
\exp(-\tau \hat{V} )
\exp \left [ -\frac{\tau (\hat{K}-\mu \hat{N})}{2}\right ] +{\cal{O}}(\tau^3),
\end{eqnarray}
where $\hat{K}$ is the kinetic energy operator, whose dispersion relation, for momenta smaller than the cut-off, is given by $\varepsilon_{\bf k}=\hbar^2 k^2/2m$. It is important to note that, because we have used the expansion up to ${\cal{O}}(\tau^{3})$, when calculating the partition function this becomes ${\cal{O}}(\tau^{2})$. Indeed, the statistical weight involves a product of $N_\tau$ factors and is given by the following expression:
\begin{equation}
\label{GibbsWeight}
\exp [-\beta (\hat{H}-\mu \hat{N})]=\exp \left [-\frac{\tau (\hat{K}-\mu \hat{N})}{2}\right ]
\left ( \prod_{j=1}^{N_\tau}\exp [-\tau \hat{V} ] \exp [-\tau (\hat{K}-\mu \hat{N})] \right ) 
\exp \left [+\frac{\tau (\hat{K}-\mu \hat{N})}{2}\right ] + {\cal{O}}(\tau^2)
\end{equation}
Notice also that this approach does not depend on the choice of dispersion relation. One may choose to implement a discrete derivative for the kinetic energy (which corresponds to the Hubbard model) based on the second difference formula:
\begin{equation}
\delta^2 f(x) = \frac{f(x+l) + f(x-l) -2f(x)}{l^2}
\end{equation}
This results in a dispersion law $\varepsilon_{k} \sim \sin^2(k_xl/2)+\sin^2(k_yl/2)+\sin^2(k_zl/2)$, where $l$ is the lattice spacing. Notice that this dispersion relation is not spherically symmetric in momentum space. In that case the energy of the higher momentum states differs noticeably from the continuum case. 
\begin{figure}[htb]
\includegraphics[scale=0.55]{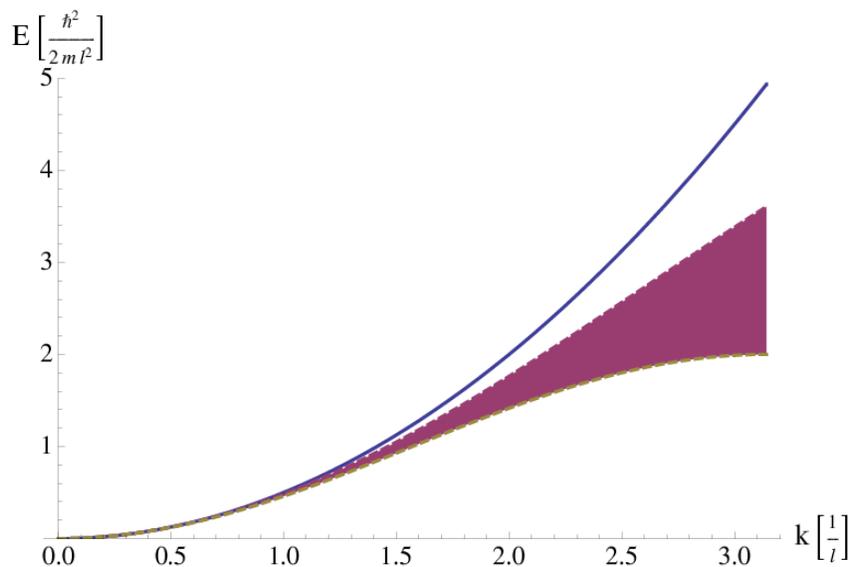}
\caption{\label{DispersionRelation} (Color online) The solid blue line shows the dispersion relation used in this work and the dashed lines and purple area result from a lowest order second difference discrete derivative (see text for discussion). The units in this plot are set by the lattice spacing: the wavevector k is in units of $1/l$ and the energy in units of $\hbar^2/2ml^2$.}
\end{figure}
Indeed, as we show in Fig.~\ref{DispersionRelation}, the dispersion relation that results from the second difference formula deviates significantly from $k^2$ behavior already at $k\simeq {\pi}/{2l}$ and as a result the number of physically meaningful available states is only $\simeq 1/2^3$ of the whole phase space. Furthermore, the deviation from the $k^2$ law depends on the angle in momentum space, and as a result the discrete derivative formula sweeps the shaded area in Fig.~\ref{DispersionRelation}. This is not the case with our choice of $\varepsilon_{\bf k}$ as we shall consider the kinetic energy operator in momentum space. According to our experience this indeed minimizes the discretization (high momentum) errors. 

The interaction factor can be represented using a discrete Hubbard-Stratonovich transformation, 
similar to the one in Ref.~\cite{hirsch}:
\begin{eqnarray}
\exp [ g \tau  \hat{n}_\uparrow ({\bf r}) \hat{n}_\downarrow ({\bf r}) ]=
\frac{1}{2}\sum _{\sigma ({\bf r},\tau_j)=\pm 1} 
[1 + A \sigma({\bf r},\tau_j)\hat{n}_\uparrow   ({\bf r}) ]
[1 + A \sigma({\bf r},\tau_j)\hat{n}_\downarrow ({\bf r}) ].
\end{eqnarray}
where $A=\sqrt{ \exp(g\tau)-1}$, $\tau_j$ labels the location on the imaginary time axis, where 
$j=1,...,N_{\tau}$, and
$\sigma({\bf r},\tau_j)$ is a field that can take values $\pm 1$ at each point on the
spacetime lattice (the name of this field should not to be confused with the spin variable, conventionally also called $\sigma$). This identity can be proven simply by evaluating both sides at $\hat{n}_{\{\uparrow,\downarrow \}} ({\bf r}) = 0,1$. This discrete Hubbard-Stratonovich transformation is sensible only for $A<1$, which means that the imaginary time step which cannot exceed $g^{-1}\log 2$.
In practice however the actual value of $\tau$ has to be much smaller due to the finite size of the $\tau$ step in formula (\ref{GibbsWeight}) (see also next section). The fact that $\sigma$ takes only discrete values is a consequence of Fermi-Dirac statistics, and 
results in a discrete configuration space for the field $\sigma$. This is the main advantage for numerical applications as compared to the case of a continuous Hubbard-Stratonovich field \cite{hirsch}.

Taking all this into account, the grand canonical partition function becomes 
\begin{equation}
Z(\beta,\mu,V) = {\mathrm{Tr}} \left \{ \exp [-\beta (\hat{H}-\mu \hat{N})] \right \} =
\int \prod_{{\bf r}\tau_j}{\cal{D}}\sigma({\bf r},\tau_j){\mathrm{Tr}}\ {\cal{\hat{U}}}(\{ \sigma \}).
\end{equation}
where we define
\begin{equation}
\label{FactorizedU}
{\cal{\hat{U}}}(\{ \sigma \})=\prod_{j=1}^{N_{\tau}} {\cal{\hat{W}}}_{j}(\{ \sigma \})
\end{equation}
and
\begin{equation}
\label{FactorizedW}
{\cal{\hat{W}}}_{j}(\{ \sigma \})=\exp \left [-\frac{\tau (\hat{K}-\mu \hat{N})}{2}\right ]\left (
\prod_{\bf i} [1 + A \sigma({\bf r},\tau_j)\hat{n}_\uparrow   ({\bf r}) ]
[1 + A \sigma({\bf r},\tau_j)\hat{n}_\downarrow ({\bf r}) ] \right )
\exp \left [-\frac{\tau (\hat{K}-\mu \hat{N})}{2}\right ].
\end{equation}
Since $\sigma$ is discrete, the integration is in fact a summation:
\begin{eqnarray}
\int \prod_{{\bf r}\tau_j}{\cal{D}}\sigma({\bf r},\tau_j) &=& \sum_{\{\sigma\}}
=\frac{1}{2^{N_{s}^{3}N_{\tau}}}
\sum_{\{\sigma({\bf r},\tau_1)\}=\pm 1}\sum_{\{\sigma({\bf r},\tau_2)\}=\pm 1}...
\sum_{\{\sigma({\bf r},\tau_{N_{\tau}})\}=\pm 1}
\end{eqnarray}
where
\begin{eqnarray}
\sum_{\{\sigma({\bf r},\tau_j)\}=\pm 1}&=&\sum_{\sigma((1,0,0),\tau_j)=\pm 1}
\sum_{\sigma((2,0,0),\tau_j)=\pm 1}...\sum_{\sigma((N_{s},N_{s},N_{s}),\tau_j)=\pm 1},
\end{eqnarray}

In a shorthand notation we will write
\begin{eqnarray}
{\cal{\hat{U}}}(\{ \sigma \})&=&{\mathrm{T}}_\tau\exp  
\left \{ -\int d \tau[\hat{h}(\{\sigma\})-\mu \hat{N}]\right \} \nonumber ,
\end{eqnarray}
where ${\mathrm{T}}_\tau$ stands for an imaginary time ordering operator and
$\hat{h}(\{\sigma\})$ is a resulting $\sigma$-dependent one-body Hamiltonian.  
It is crucial to note that ${\cal{\hat{U}}}(\{ \sigma \})$ can be expressed as a product
of two operators which describe the imaginary time evolution of the spin-up and spin-down fermions:
\begin{eqnarray} \label{decomp}
{\cal{\hat{U}}}(\{ \sigma \})&=&
{\cal{\hat{U}}}_{\downarrow}(\{ \sigma \}){\cal{\hat{U}}}_{\uparrow}(\{ \sigma \}), 
\\
{\cal{\hat{U}}}_{\downarrow}(\{ \sigma \})&=&\prod_{j=1}^{N_{\tau}} 
{\cal{\hat{W}}}_{j\downarrow}(\{ \sigma \}),
\mbox{\hspace{1cm}}
{\cal{\hat{U}}}_{\uparrow}(\{ \sigma \})=\prod_{j=1}^{N_{\tau}} {\cal{\hat{W}}}_{j\uparrow}(\{ \sigma \}). \nonumber
\end{eqnarray}
The operators for spin-up and spin-down are identical for the case in which the chemical potential is the same for each spin: $\mu_\uparrow = \mu_\downarrow = \mu$ (which is the case we consider in this work).

The expectation values of operators take the form
\begin{equation} \label{eq:ham-T}
\!\!\!\!\!\!\!\!\!\!\!\!
O(\beta,\mu,V) = 
\frac{{\mathrm{Tr}} \; \left \{ \hat{O}\exp [-\beta (\hat{H}-\mu \hat{N})] \right \}}
{Z(\beta,\mu,V)}=
\int \frac{\prod_{{\bf i}j}{\cal{D}}\sigma({\bf r},\tau_j)
{\mathrm{Tr}}\;{\cal{\hat{U}}}(\{ \sigma \})}{Z(\beta,\mu,V)}
\; \frac{ {\mathrm{Tr}}\; \hat{O}{\cal{\hat{U}}}(\{ \sigma \})}
{{\mathrm{Tr}}\; {\cal{\hat{U}}}(\{ \sigma \})}
\end{equation}
where we have introduced ${{\mathrm{Tr}}\; {\cal{\hat{U}}}(\{ \sigma \})}$ for convenience: in the numerator it represents the probability measure used in our simulations (see below), and in the denominator it serves the purpose of moderating the variations of ${\mathrm{Tr}}\; \hat{O}{\cal{\hat{U}}}(\{ \sigma \})$ as a function of the auxiliary field $\sigma$.

All of the above traces over Fock space acquire very simple forms \cite{svd,yoram} (see next section), and can be easily evaluated. In particular, ${\mathrm{Tr}}\;{\cal{\hat{U}}}(\{ \sigma \})$ can be written as
\begin{equation} 
{\mathrm{Tr}}\; {\cal{\hat{U}}}(\{ \sigma \})= \det [1 + {\cal{U}}(\{ \sigma \})] =
\det [1 + {\cal{U}}_{\downarrow}(\{ \sigma \})] \det [1 + {\cal{U}}_{\uparrow}(\{ \sigma \})],
\end{equation}
where ${\cal U}$ (without the hat) is the representation of $\hat {\cal U}$ in the 
single-particle Hilbert space. 
The second equality is a result of the decomposition (\ref{decomp}).
This identity is easy to prove by expanding both 
sides, taking into account that ${\cal{U}}$ is a product of exponentials of
1-body operators. In the case considered in this work the chemical potential is the same for spin-up and spin-down fermions, so it follows that
$\det [1 + {\cal{U}}_{\downarrow}(\{ \sigma \})] = \det [1 + {\cal{U}}_{\uparrow}(\{ \sigma \})]$.
This implies that ${\mathrm{Tr}}\; {\cal{\hat{U}}}(\{ \sigma \})$ is
positive, i.e., that there is no fermion sign problem for this system. Indeed, this allows
to define a positive definite probability measure:
\begin{equation}  \label{eq:measure}
P(\{ \sigma \}) = \frac{{\mathrm{Tr}}\; {\cal{\hat{U}}}(\{ \sigma \})}{Z(\beta,\mu,V)} = \frac{\{\det [1 + {\cal{U}}_{\uparrow}(\{ \sigma \})]\}^2}{Z(\beta,\mu,V)} =
\frac{1}{Z(\beta,\mu,V)} 
\exp(2~\text{tr}\left(\log [1 + {\cal{U}}_{\uparrow}(\{ \sigma \})]\right))
\end{equation}
where the exponent in the last equation defines the negative of the so-called effective action.

The many-fermion problem is thus reduced to an auxiliary field 
Quantum Monte Carlo problem, to which the standard Metropolis algorithm 
can be applied, using Eq.~(\ref{eq:measure}) as a probability measure. 
Before moving on to the details of our Monte Carlo algorithm, we briefly
discuss the expressions used to compute a few specific thermal averages.

\subsection{Calculation of observables}

Let us consider the one body operator
\begin{equation}
\hat{O}=\sum_{s,t=\downarrow,\uparrow}
\int d^3 {\mathbf r_{1}} d^3 {\mathbf r_{2}}\hat{\psi}_{s}^{+}(\mathbf{r_{1}})
O_{st}({\mathbf r_{1}},{\mathbf r_{2}})\hat{\psi}_{t}(\mathbf{r_{2}})
\end{equation}
From Eq.~(\ref{eq:ham-T}) it follows that
\begin{equation}
\langle \hat{O}\rangle  = \sum_{\{\sigma\}}P(\{\sigma\})\frac{ {\mathrm{Tr}}\; 
\hat{O}{\cal{\hat{U}}}(\{ \sigma \})}
{{\mathrm{Tr}}\; {\cal{\hat{U}}}(\{ \sigma \})}=
\sum_{\{\sigma\}}P(\{\sigma\})\frac{ {\mathrm{Tr}}\; \hat{O}{\cal{\hat{U}}}(\{ \sigma \})}
{\det [1+{\cal{U}}(\{ \sigma \})]}.
\end{equation}
The calculation of the last term requires the evaluation of
\begin{equation}
{\mathrm{Tr}}\left [ 
\hat{\psi}_{s}^{+}(\mathbf{r_{1}})\hat{\psi}_{t}(\mathbf{r_{2}}){\cal{\hat{U}}}(\{ \sigma \}) \right ]
= \delta_{st}\det[1+{\cal{U}}(\{ \sigma \})]^2 n_{s}({\bf r_{1}},{\bf r_{2}},\{ \sigma\})
\end{equation}
where $s,t$ denote spin ($\uparrow$ or $\downarrow$), and
\begin{equation}
n_{s}({\bf r_{1}},{\bf r_{2}},\{ \sigma\})= 
 \sum_{{\bf k_1},{\bf k_2}\le k_c}
\varphi_{\bf k_1}({\bf r_{1}}) 
\left [ \frac{  {\cal{U}}_{s}(\{ \sigma \})  }{  1+{\cal{U}}_{s}(\{ \sigma \})   }
\right ] _{ {\bf k_1},{\bf k_2} } \varphi_{\bf k_2}^*({\bf r_{2}})
\end{equation}
Here $\varphi_{\bf k}({\bf r})=\exp(i{\bf k}\cdot{\bf r})/L^{3/2}$ are the single-particle
orbitals on the lattice with periodic boundary conditions.
This holds for any 1-body operator $\hat O$, if $\cal{U}$ is a product of exponentials 
of 1-body operators, as is the case once the Hubbard-Stratonovich 
transformation is performed. 
It is then obvious that the momentum representation of the one-body density matrix has the form
\begin{equation}
n_{s}({\bf k_1},{\bf k_2},\{ \sigma\})= 
 \left [ \frac{  {\cal{U}}_{s}(\{ \sigma \})  }{  1+{\cal{U}}_{s}(\{ \sigma \})   }
\right ] _{ {\bf k_1},{\bf k_2} }
\end{equation}
which, for a noninteracting homogeneous Fermi gas, is diagonal and equal to the 
occupation number probability $1/(\exp [\beta (\varepsilon_{\bf k}-\mu)] + 1)$ of a state with the energy $\varepsilon_{\bf k}=\displaystyle{\frac{\hbar^{2}k^{2}}{2m}}$. 

Summarizing, the expectation value of any one-body operator may be calculated by summing 
over samples of the auxiliary field $\sigma({\bf r},\tau_j)$:
\begin{equation}
\langle \hat{O}\rangle  = \int\prod_{{\bf r}\tau_j}{\cal{D}}\sigma({\bf r},\tau_j) P(\{\sigma \})
\sum_{{\bf r_1},{\bf r_2}}\sum_{s=\uparrow,\downarrow}
O_{ss}({\mathbf r_{1}},{\mathbf r_{2}})n_{s}({\bf r_{1}},{\bf r_{2}},\{ \sigma\})
\end{equation}

In particular the kinetic energy can be calculated according to:
\begin{equation}
\langle \hat{K}\rangle = 
\int \frac{\prod_{{\bf r}\tau_j}{\cal{D}}\sigma({\bf r},\tau_j)
{\mathrm{Tr}}\;{\cal{U}}(\{ \sigma \})}{Z(\beta,\mu,V)}
\; \frac{ {\mathrm{Tr}}\; \hat{K}{\cal{U}}(\{ \sigma \})}
{{\mathrm{Tr}}\; {\cal{U}}(\{ \sigma \})} =
\int \prod_{{\bf r}\tau_j}{\cal{D}}\sigma({\bf r},\tau_j)P(\{\sigma \})
\sum_{{\bf k}}^{k\le k_{c}}\sum_{s=\uparrow,\downarrow}
\left [ n_{s} ({\bf k},{\bf k},\{ \sigma\}) \frac{\bf \hbar^{2}k ^2}{2m}
\right ]
\end{equation}

Analogously, for a generic two-body operator:
\begin{equation}
\hat{O}=\sum_{s,t,u,v=\downarrow,\uparrow}
\int d^3 {\mathbf r_{1}'} d^3 {\mathbf r_{2}'} d^3 {\mathbf r_{1}}d^3 {\mathbf r_{2}}
\hat{\psi}_{s}^{+}(\mathbf{r_{1}'})\hat{\psi}_{t}^{+}(\mathbf{r_{2}'})
O_{stuv}({\mathbf r_{1}'},{\mathbf r_{2}'},{\mathbf r_{1}},{\mathbf r_{2}})
\hat{\psi}_{v}(\mathbf{r_{2}})\hat{\psi}_{u}(\mathbf{r_{1}})
\end{equation}
in order to calculate $\langle\hat{O}\rangle$ one needs to evaluate expression
\begin{eqnarray}
&{\mathrm{Tr}}&\left [ 
\hat{\psi}_{s}^{+}(\mathbf{r_{1}'})\hat{\psi}_{t}^{+}(\mathbf{r_{2}'})
\hat{\psi}_{v}(\mathbf{r_{2}})\hat{\psi}_{u}(\mathbf{r_{1}})
{\cal{\hat{U}}}(\{ \sigma \}) \right ]= \nonumber \\
&=&\left(\det[1+{\cal{U}}(\{ \sigma \})]\right )^2 \left (
\delta_{su}\delta_{tv}
 n_{s}({\bf r_{1}'},{\bf r_{1}},\{ \sigma\}) n_{t}({\bf r_{2}'},{\bf r_{2}},\{ \sigma\})-
\delta_{sv}\delta_{tu}n_{s}({\bf r_{1}'},{\bf r_{2}},\{ \sigma\}) n_{t}({\bf r_{2}'},{\bf r_{1}},\{ \sigma\})
\right ).
\end{eqnarray}
Hence, for the expectation value of the two body operator we get
\begin{eqnarray}
\langle\hat{O}\rangle=
\int\prod_{{\bf r}\tau_j}{\cal{D}}\sigma({\bf r},\tau_j) P(\{\sigma \}) 
\sum_{{\bf r_{1}'},{\bf r_{2}'},{\bf r_1},{\bf r_2}}\sum_{s,t=\uparrow,\downarrow}
\Bigg [
&& O_{stst}({\mathbf r_{1}'},{\mathbf r_{2}'},{\mathbf r_{1}},{\mathbf r_{2}})
n_{s}({\bf r_{1}'},{\bf r_{1}},\{ \sigma\}) n_{t}({\bf r_{2}'},{\bf r_{2}},\{ \sigma\}) - \nonumber \\
&&O_{stts}({\mathbf r_{1}'},{\mathbf r_{2}'},{\mathbf r_{1}},{\mathbf r_{2}})
n_{s}({\bf r_{1}'},{\bf r_{2}},\{ \sigma\}) n_{t}({\bf r_{2}'},{\bf r_{1}},\{ \sigma\})
\Bigg ].
\end{eqnarray}
In particular, the expectation value of the interaction energy reads:
\begin{equation}
\langle\hat{V}\rangle=-g\int\prod_{{\bf r}\tau_j}{\cal{D}}\sigma({\bf r},\tau_j) P(\{\sigma \}) 
\sum_{\bf r}n_{\uparrow}({\bf r},{\bf r},\{ \sigma\}) n_{\downarrow}({\bf r},{\bf r},\{ \sigma\})
\end{equation}
It should be noted that in the symmetric system ($\mu_{\uparrow}=\mu_{\downarrow}$)
\begin{equation}
n_{\uparrow}({\bf r},{\bf r}',\{ \sigma\}) = n_{\downarrow}({\bf r},{\bf r}',\{ \sigma\}).
\end{equation}
Hence,
\begin{equation}
\langle\hat{V}\rangle=-g\int\prod_{{\bf r}\tau_j}{\cal{D}}\sigma({\bf r},\tau_j) P(\{\sigma \}) 
\sum_{\bf r} [ n_{\uparrow}({\bf r},{\bf r},\{ \sigma\}) ]^{2}
\end{equation}
It is useful to introduce the correlation function
\begin{eqnarray}
g_2({\bf r}) &=&\left (\frac{2}{N}\right )^2
\int d^3 {\mathbf r_{1}} d^3 {\mathbf r_{2}} 
\langle\psi_\uparrow^\dagger({\bf r_{1}+r})
\psi_\downarrow^\dagger({\bf r_{2}+r})\psi_\downarrow({\bf r_{2}})
\psi_\uparrow({\bf r_{1}})\rangle \; \nonumber \\ 
&=& \left (\frac{2}{N}\right )^{2}
\int\prod_{{\bf r}\tau_j}{\cal{D}}\sigma({\bf r},\tau_j) P(\{\sigma \}) \nonumber \\
&\times& \int d^3 {\mathbf r_{1}} d^3 {\mathbf r_{2}} 
n_{\uparrow}({\bf r_{1}+r},{\bf r_{1}},\{ \sigma\}) 
n_{\downarrow}({\bf r_{2}+r},{\bf r_{2}},\{ \sigma\})
\label{TBDM}
\end{eqnarray}
(where $N$ is the average particle number) 
which is normalized in such a way that for a noninteracting homogeneous Fermi gas 
$g_{2} ({\bf r})=\displaystyle{3\frac{j_{1}(k_{F}r)}{k_{F}r}}$ and $g_{2} (0)=1$.

\section{Numerical methods and computational issues}
\subsection{Metropolis Monte Carlo}

Once we have written the observables as in Eq.~(\ref{eq:ham-T}), the next step is to 
sum over all possible configurations of $\sigma({\bf r},\tau_j)$. For a lattice size $N_x^3 \times N_\tau$ (where typically $N_x = 8$ and $N_\tau \simeq 1000$), performing the sum over the $2^{N_x^3 \times N_\tau}$ points in configuration space, is an impossible task for all practical purposes. It is in these cases that a Monte Carlo (MC) approach becomes essential. By generating $\cal N$ independent samples of the field $\sigma({\bf r},\tau_j)$ with probability given by (\ref{eq:measure}), and adding up the values of the integrand at those samples, one can estimate averages of observables with $O(1/\sqrt{\cal N})$ accuracy.

The standard Metropolis algorithm was chosen to generate the samples.  At every MC step the sign of $\sigma$ was changed at randomly chosen locations on the space-time lattice. An adaptive routine increased or decreased the fraction of sites where $\sigma$ was updated, so as to maintain an average acceptance rate (over $\simeq$ 100 consecutive Metropolis steps) between 0.4 and 0.6. 

In order to compute the probability of a given $\sigma$ configuration, it is necessary to find the matrix elements of $\cal U$, which entails applying it to a complete set of single-particle wave-functions. For the latter we chose plane waves (with momenta $\hbar k\le \hbar k_c$). This choice is particularly convenient because one can compute the overlap of any given function with the whole basis of plane waves by performing a single Fast Fourier Transform (FFT) on that function.

In practice, the calculation of the matrix elements of $\cal U$ proceeds by evolving the above mentioned set of wave-functions in imaginary time, applying the $N_\tau$ operators ${\cal \hat W}_j$ in Eq.~(\ref{FactorizedU}) in sequence. The operator ${\cal \hat W}_j$ is in turn made up of kinetic energy and potential energy factors (see Eq.~(\ref{FactorizedW})), in a specific order. The application of such factors was implemented in momentum space and real space, respectively, using FFT to switch between them.

\subsection{Singular value decomposition}

\begin{figure}[htb]
\includegraphics[scale=0.35]{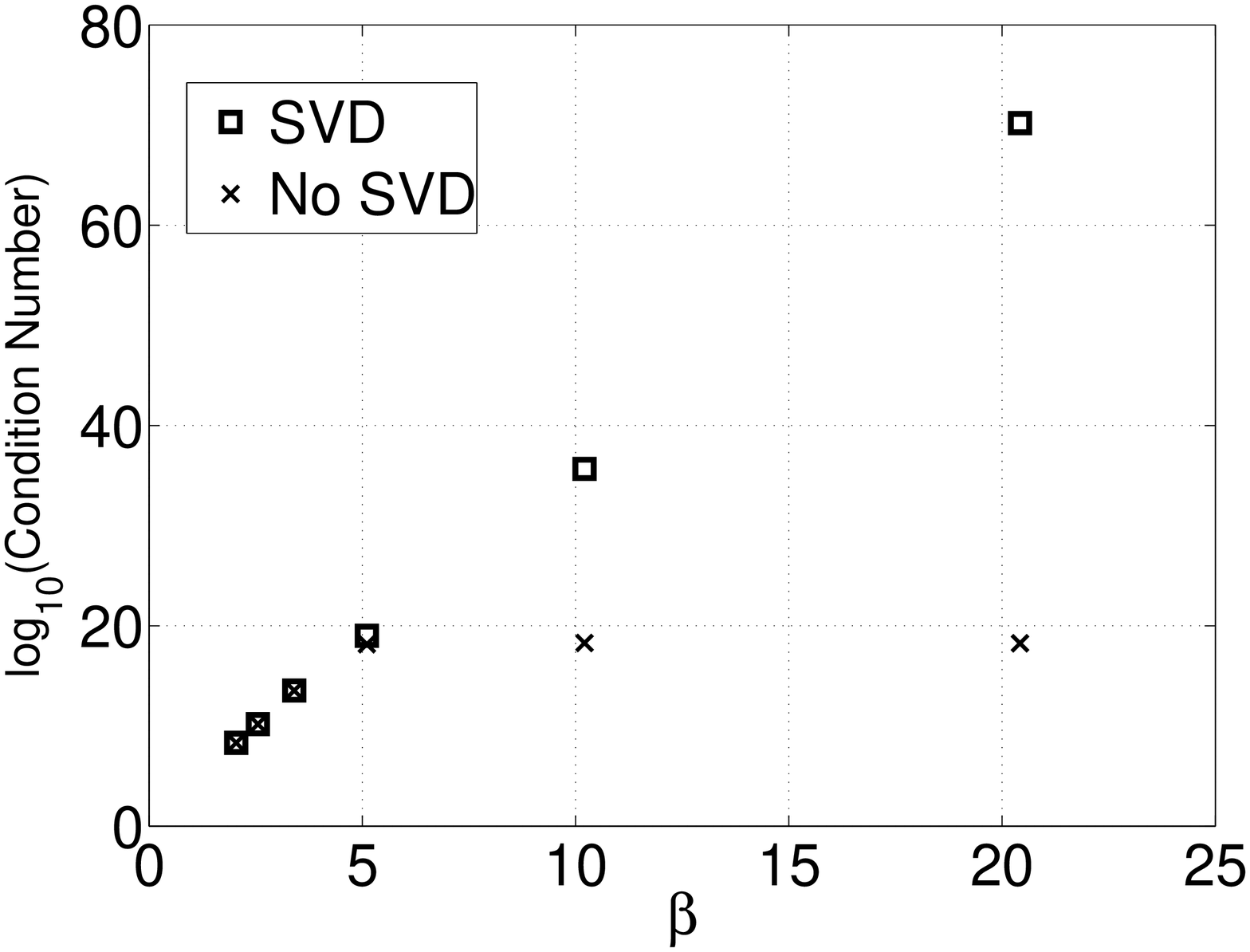}
\includegraphics[scale=0.35]{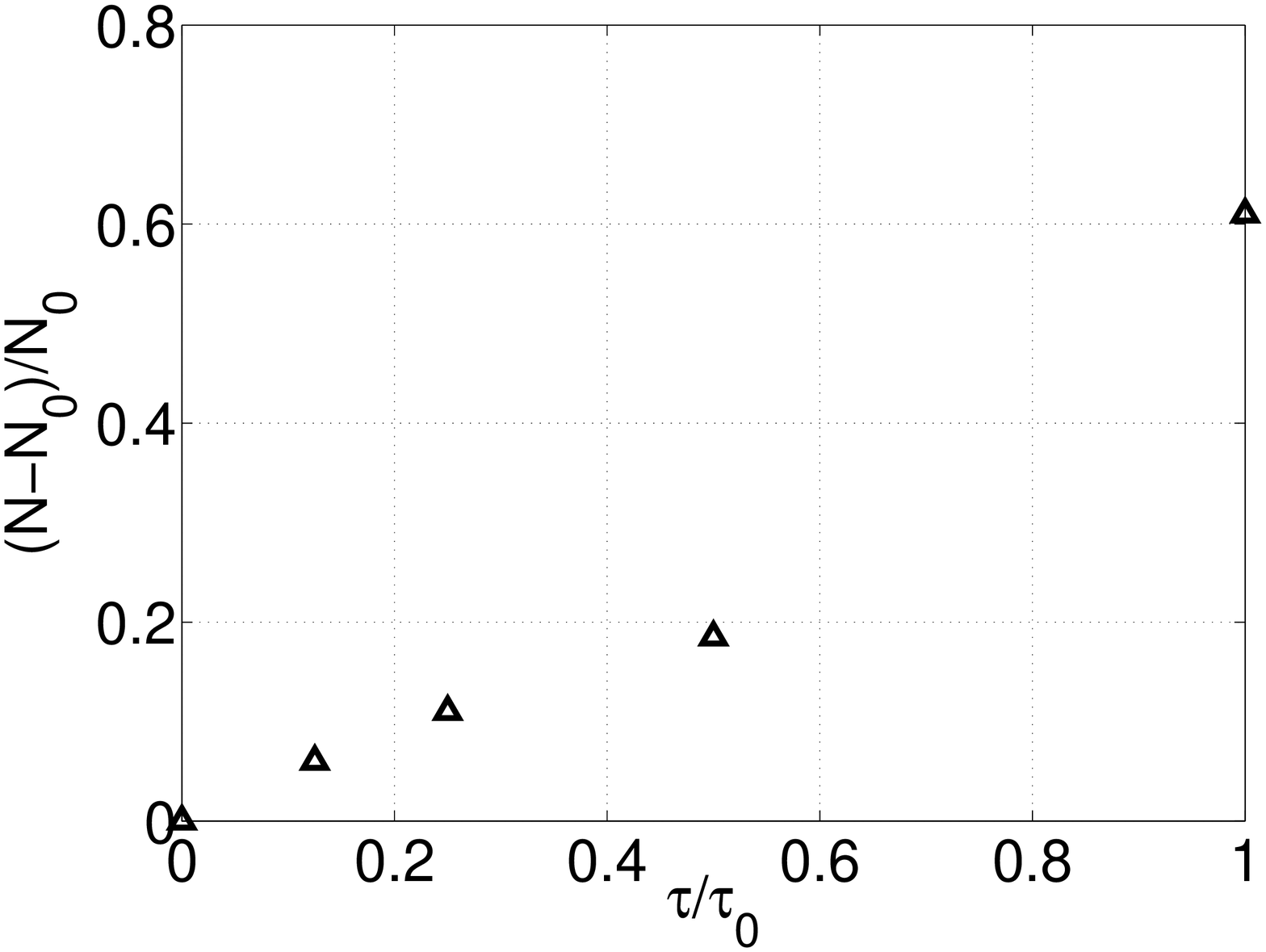}
\caption{\label{CNCT} Left Panel: Condition number of $\cal U$ as a function of 
$\beta$. Squares: with SVD. Crosses: without SVD. Right Panel: Convergence 
of simulated particle number $N$ relative to exact solution $N_0$, as a function 
of time step $\tau$ in units of $\tau_0$, defined in the text.}
\end{figure}

Matrix multiplication, especially in the form of FFT's, is ubiquitous in our algorithm. 
It is well known that matrix multiplication is numerically unstable when the matrices involved have elements that vary over a large range. This is true in our case at low $T$, with exponentially diverging scales. To avoid instabilities it is necessary to separate the scales when multiplying matrices, and the Singular Value Decomposition (SVD) technique serves such purpose. In this section we follow the same approach developed in Ref.~\cite{svd} to introduce SVD in our calculations.

Let us write the matrix $\cal U(\{ \sigma \})$ more explicitly:
\begin{equation}
{\cal{U}}(\{ \sigma \})=
\prod_{j=1}^{N_\tau} {\cal{W}}_j(\{ \sigma \}) = 
{\cal{W}}_{N_\tau} {\cal{W}}_{N_{\tau}-1} ... {\cal{W}}_2 {\cal{W}}_1  
\end{equation}
where the ${\cal{W}}_k(\{ \sigma \})$ will be $N\times N$ matrices, for a single-particle basis of dimension $N$. Let us then define 
\begin{eqnarray}
{\cal{U}}_{0} &=& 1 \\ \nonumber
{\cal{U}}_{1} &=& {\cal{W}}_{1} \\ \nonumber
{\cal{U}}_{2} &=& {\cal{W}}_{2}{\cal{W}}_{1} \\ \nonumber
.&& \\ \nonumber
.&& \\ \nonumber
.&& \\ \nonumber
{\cal{U}}_{n} &=& {\cal{W}}_{n}{\cal{W}}_{n-1}... {\cal{W}}_{1} = 
{\cal{W}}_{n}{\cal{U}}_{n-1}. \nonumber
\end{eqnarray}

To separate the scales one decomposes the matrix ${\cal{U}}_{n-1}$
before multiplying it by ${\cal{W}}_{n}$ to get ${\cal{U}}_{n}$. This process begins as follows
\begin{eqnarray}
{\cal{U}}_{0} &=& 1 \\ \nonumber
{\cal{U}}_{1} &=& {\cal{W}}_{1} = {\cal{S}}_{1}{\cal{D}}_{1}{\cal{V}}_{1} \\ \nonumber
{\cal{U}}_{2} &=& {\cal{W}}_{2}{\cal{W}}_{1} = ({\cal{W}}_{2}{\cal{S}}_{1}{\cal{D}}_{1}){\cal{V}}_{1} = {\cal{S}}_{2}{\cal{D}}_{2}{\cal{V}}_{2}{\cal{V}}_{1} \nonumber
\end{eqnarray}
where ${\cal{S}}_{1}$ and ${\cal{V}}_{1}$ are orthogonal matrices (not necessarily 
inverse of each other), and ${\cal{D}}_{1}$ is a diagonal positive matrix containing the
singular values of ${\cal{U}}_{1}$. The idea is that the actual multiplication should be done by first computing the factor in parenthesis in the last equation. 
This factor is then decomposed into ${\cal{S}}_{2}{\cal{D}}_{2}{\cal{V}}_{2}$, in preparation for the multiplication by ${\cal{W}}_{3}$, and so on. 
A generic step in this process looks like this:
\begin{equation}
{\cal{U}}_{n} = {\cal{W}}_{n}{\cal{U}}_{n-1} = 
{\cal{W}}_{n}{\cal{S}}_{n-1}{\cal{D}}_{n-1}{\cal{V}}_{n-1}{\cal{V}}_{n-2}...{\cal{V}}_{1}
\end{equation}
so in the end
\begin{equation}
{\cal{U}}_{N_\tau} = {\cal{U}}(\{\sigma\})= 
{\cal{S}}_{N_\tau}{\cal{D}}_{N_\tau}{\cal{V}}_{N_\tau}{\cal{V}}_{N_\tau-1}...{\cal{V}}_{1}=
{\cal{S}}{\cal{D}}{\cal{V}}
\end{equation}
where we have decomposed the full product in the last step.
Calculation of the determinant, and therefore of the probability measure, 
becomes straightforward if we perform one more
SVD, in the following chain of identities:
\begin{equation}
\det(1+ {\cal{U}}(\{\sigma\})) = \det(1+{\cal{S}}{\cal{D}}{\cal{V}}) =
\det({\cal{S}}({\cal{S}}^\dagger{\cal{V}}^\dagger+{\cal{D}}){\cal{V}}) =
\det({\cal{S}} \  \tilde{\cal{S}} \tilde{\cal{D}} \tilde{\cal{V}}\ {\cal{V}}) =
\det({\cal{S}}\tilde{\cal{S}})\det(\tilde{\cal{D}})\det( \tilde{\cal{V}}{\cal{V}})
\end{equation}

For equal densities (symmetric case), where we need this determinant squared, we only care about the factor in the middle of the last expression, since the other two are equal to 1 in magnitude. Indeed, in that case we can write the probability measure as
\begin{equation}
P(\{\sigma\}) = \exp\left(\sum^M_{i=1} \log \tilde d_{i}\right)
\end{equation}
where $\tilde d_{i} > 0$ are the elements in the diagonal of $\tilde{\mathcal D}$, and $M$ is the dimension of the single particle Hilbert space.

The SVD decomposition is useful when evaluating occupation numbers. Indeed,
\begin{equation}
\frac{{\cal U}(\{\sigma\})}{1+{\cal U}(\{\sigma\})}=1-\frac{1}{1+{\cal U}(\{\sigma\})}
=1-
{\cal{V}}^{\dagger}\tilde{\cal{V}}^{\dagger}\tilde{\cal{D}}^{-1}\tilde{\cal{S}}^{\dagger}{\cal{S}}^{\dagger}
\end{equation}
which is very easy to compute since the matrix $\tilde{\cal{D}}$ is diagonal.

In Fig.~\ref{CNCT} (left panel) we show the behavior of the condition number (defined as 
the ratio of largest to smallest eigenvalue) of the matrix ${\cal U}_{N_\tau}$ in the single-particle Hilbert 
space. The number of SVD's required to stabilize the calculation grows as we increase $\beta$. 
If no SVD's are used, the condition number saturates, indicating loss of information due to poor separation 
of scales in matrix multiplication. In our calculations we have made limited use of SVD, ranging from 2 decompositions at the highest $T$ to 8 decompositions at low $T$'s.

\subsection{Tests and cross-checks
\label{TestsCrossChecks}}

In order to verify the correctness of our code we performed several tests. As a first check the thermodynamics of a free gas was reproduced when setting $g = 0$. This is an elementary
test, as in this case the MC part of the algorithm is superfluous.

To check our results at $g \neq 0$ we diagonalized the Hamiltonian exactly, restricting the 
phase space to the lowest 7 single-particle momentum states. This entails constructing all of 
the $2^{14}$ states ($2^7$ for each spin) and computing and block diagonalizing the 
Hamiltonian (which comes in blocks identified by fixed particle numbers 
$(N_{\uparrow},N_{\downarrow}))$. An average desktop computer can complete the whole task in about 5 minutes. This test provided an estimate for the size of the step in the imaginary time direction. In Fig.~\ref{CNCT} (right panel) we plot the difference between the simulated number of particles $N$ and the exact value $N_0$, as a function of the time step, in units of $\tau_0 = \ln 2 /g$, with all the other parameters ($T, \mu$, etc) fixed. We conclude from this data that it is safe to take $\tau = \tau_0/10$, as the relative error falls below 10\%.

In the unitary limit universality implies thermodynamic relations that provide self-consistency checks. For instance, it can be proven that if $E = 3/5 \varepsilon_F N \xi(T/\varepsilon_F)$ then
\begin{equation}
\label{ConsistencyCheck}
\frac{\mu}{\varepsilon_F} = \xi \left(\frac{T}{\varepsilon_F} \right) - 
\frac{3}{5} \frac{T}{\varepsilon_F}  \int^{T/\varepsilon_F}_0 dy \frac{\xi'(y)}{y}.
\end{equation}
In our simulations $\mu$, $T$ and $V$ are input parameters, while $E$ and $\varepsilon_F$ are part of the output. We have checked that the above relation is satisfied by our data.

\subsection{Density dependence and the role of periodic boundary conditions and the high-momentum cutoff }
\label{sec:densitydependence}
\begin{figure}[htb]
\includegraphics[scale=0.45]{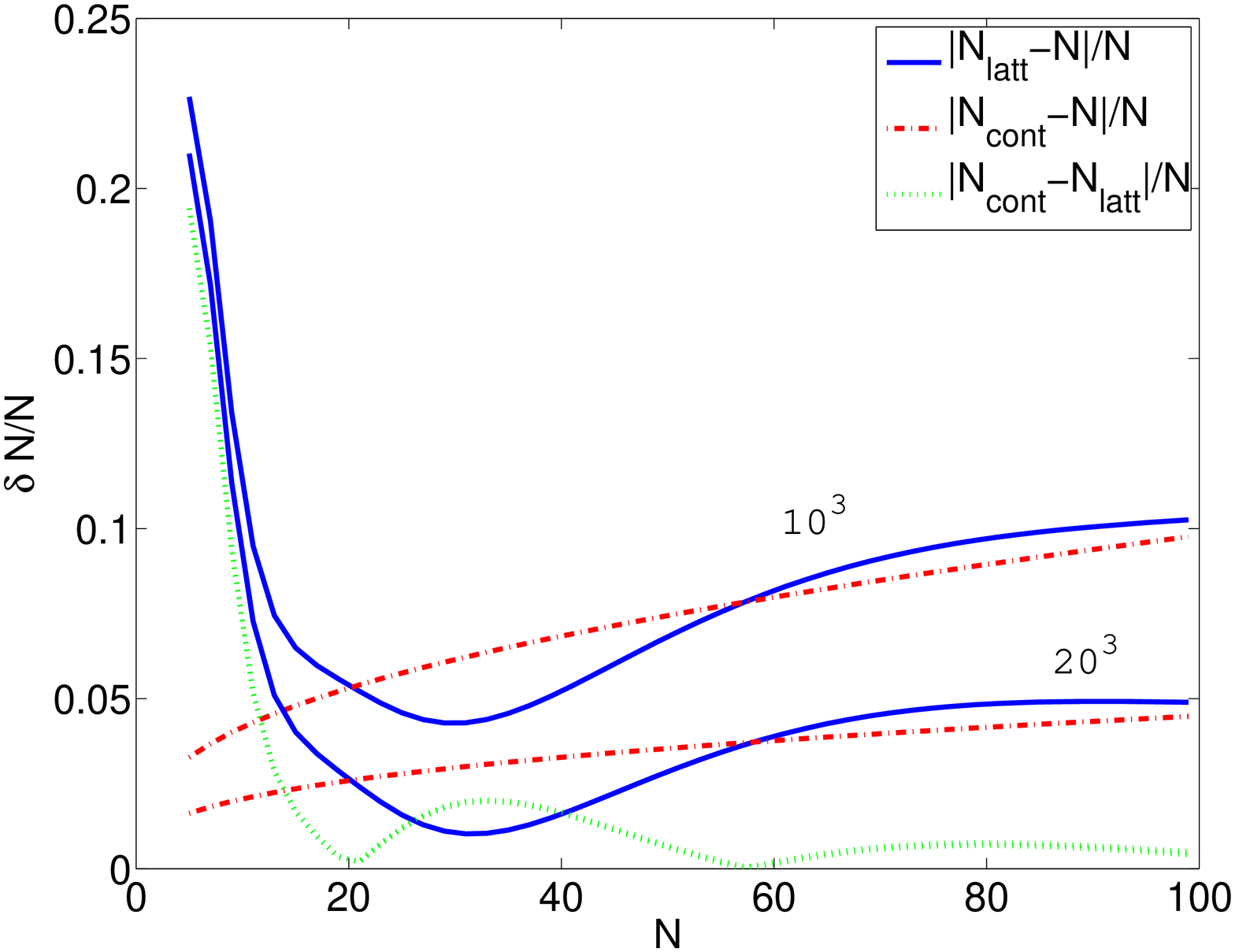}
\caption{\label{fig:boundary_conditions}. (Color online) The error in particle number obtained in a simulation in a box with periodic boundary conditions and upper momentum cutoff $|N_{latt}-N|/N$, solid line (blue), compared with the error obtained in a truly continuum model with no periodic boundary conditions imposed but only an upper momentum cutoff $|N_{cont}-N|/N$, dash-dotted line (red), and the "numerical" apparent error 
$|N_{latt}-N_{cont}|/N$, dotted line (green). We display here two cases, of simulations in a box with lattice sizes $10^3$ and with $20^3$. In the case of the quantity $|N_{latt}-N_{cont}|/N$ we see no lattice size dependence. 
  }
\end{figure}

An MC simulation of a Fermi gas in the unitary regime makes sense only if the following conditions are satisfied:
\begin{equation}\label{eq:k_cut}
\Lambda_0=\frac{2\pi}{L} \ll  k_F \ll k_c=\frac{\pi}{l}.
\end{equation}
The size of the box $L$ defines the lowest momentum scale $\hbar\Lambda_0=2\pi\hbar/L$ one can resolve in such a simulation, while the lattice constant $l$ defines the smallest inter-particle separation accessible.  To better appreciate how the restriction (\ref{eq:k_cut}) affects the simulations we present in Fig. \ref{fig:boundary_conditions} the errors one incurs due to both the upper momentum cutoff, needed to regularize the two-body interactions, and the use of boundary conditions. We calculate the total particle number in a box with periodic boundary conditions in three different ways:
\begin{eqnarray}
N = L^3 \int \frac{d^3k}{(2\pi)^3} \left [ 1 - \frac{\epsilon (k) +U -\mu}{\sqrt{(\epsilon (k) +U -\mu)^2+\Delta^2}} \right ] ,\\
N_{cont} = L^3 \int_{k\le k_c} \frac{d^3k}{(2\pi)^3}  
   \left [ 1 - \frac{\epsilon (k) +U -\mu}{ \sqrt{(\epsilon (k) +U -\mu)^2+\Delta^2} } \right ] ,\\
N_{latt} = \sum _{ {\bf k}} ^{ k\le k_c} \left [ 1 - \frac{ \epsilon (k) +U -\mu }{ \sqrt{(\epsilon (k) +U -\mu)^2+\Delta^2} } \right ],
\end{eqnarray}
where $\epsilon (k)$ is the single-particle kinetic energy, ${\bf k}=2\pi/L(n_x,n_y,n_z)$ and $n_{x,y,z}$ are integers, and 
using parameters characteristic for the unitary Fermi gas at zero temperature, see Eq. (\ref{eq:ep}) and Ref. \cite{slda}. Thus $N$ is the exact particle number in such a box if no restrictions on momenta were imposed, $N_{cont}$ would be the actual particle number only if all momenta smaller than the upper cutoff momentum $\hbar k_c=\pi\hbar/l$ would be taken into account, and $N_{latt}$ is the particle number one would obtain if the simulation were performed in a box with periodic boundary conditions and an upper cutoff momentum $\hbar k_c$. The quantity $|N-N_{cont}|/N$ is thus a measure of the error introduced by a finite cutoff momentum $\hbar k_c$ alone.
It is debatable whether one has to take into account in $N_{cont}$ the contributions from momenta larger than the cutoff momentum $\hbar k_c$, since the physics at those momenta is not simulated correctly anyway. In a box with periodic boundary conditions by default one can access only a finite set of discrete momenta. As it is clear from the figure, imposing periodic boundary conditions alone leads to very large errors especially in the case of small densities, when the Fermi momentum $\hbar k_F=\hbar (3\pi^2N/L^3)^{1/3}$ becomes smaller than the momentum $\hbar \Lambda_0=2\pi\hbar/L$. It is important to notice as well that the magnitude of these errors are independent of the presence or absence of the upper cutoff momentum $\hbar k_c$. A rule of thumb would suggest that one has to have at least ten particles in a box in order to keep the errors at an acceptable level.  This is relatively easy to understand physically, as in a box with periodic boundary conditions the lowest single-particle levels are: one level with zero energy, six levels with energy 
$\hbar^2\Lambda_0^2/2m$, and twelve levels with energy $2\hbar^2\Lambda_0^2/2m$, not counting spin degeneracy. 

The quantity $k_FL/2\pi$ is basically the ratio between half of the box size ($L/2$) and the average inter-particle separation ($\approx \pi/k_F$), and in our simulations we were able to probe $k_FL/2\pi\approx 2.5$. In Ref. \cite{newdata} this ratio was increased to about 5, i.e. the system was significantly more dilute. Notice however, the errors incurred in a calculation in a box with periodic boundary conditions and very small particle number (and thus very low density) become unacceptably large, as demonstrated in Fig. \ref{fig:boundary_conditions}.  

With the densities used in this work, we find excellent agreement between our results and GFMC and DMC results \cite{carlson,chang,giorgini,stefano,CarlsonReddy,GezerlisCarlson} for essentially all of the quantities of interest. All these calculations however were performed typically for particle numbers between 10 \cite{carlson,chang} and less than 66 \cite{carlson,chang,giorgini,stefano,CarlsonReddy,GezerlisCarlson}, where the errors arising from imposing periodic boundary conditions are minimal. Only the long range universal critical behavior of the two-body density matrix discussed in Section IV ($G(r) \propto r^{-(1+\eta)}$) requires very low filling factors, but in very large box sizes, and in a very narrow temperature interval around the transition.

\section{Results in the unitary limit}

\begin{figure}[htb]
\centering
\includegraphics[scale=0.48]{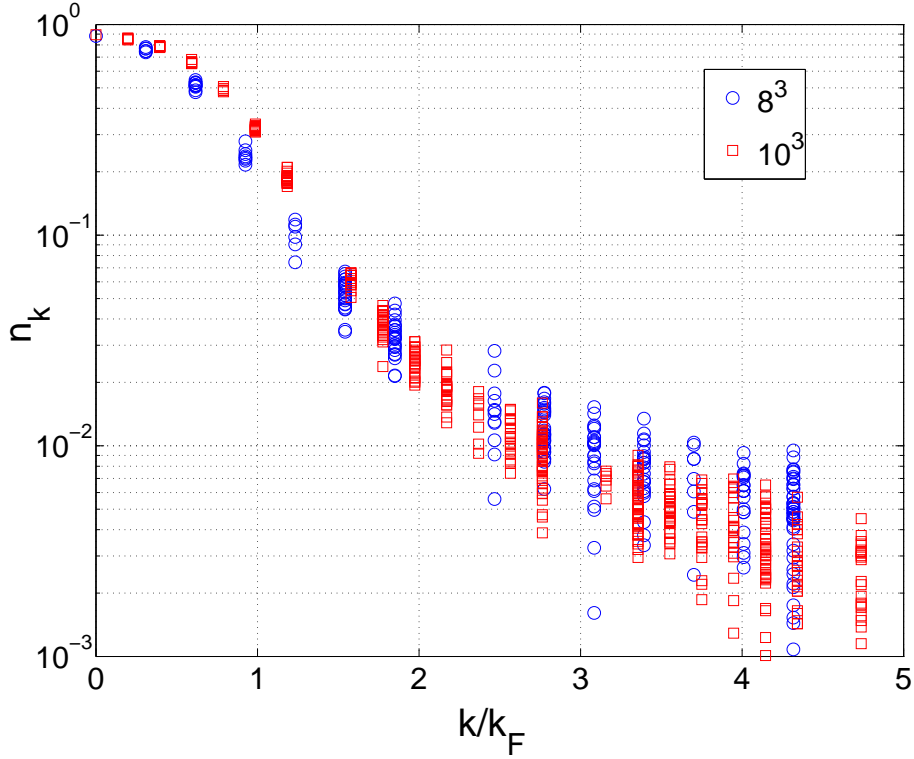}
\caption{ \label{occup} (Color online)  Occupation probability in y-log scale at intermediate temperatures, $T=0.20 \varepsilon_F$, in blue circles for $8^3$ and red squares for $10^3$.}
\end{figure}

The results of our simulations for lattices ranging from $8^3\times 1732$ (at low $T$) 
to $8^3\times 257$ (at high $T$) and for $2\cdots 20\times 10^5$ Monte Carlo 
samples (after thermalization) are shown in Fig.~\ref{energyentropy}. 
The imaginary time step was chosen as $\tau = \min (ml^2/15\pi^2\hbar^2, \ln 2/10g)$, The first bound comes from the inverse of the highest single-particle kinetic energy available on the lattice, namely $E_{k,max} = 3 \hbar^2 \pi^2/2ml^2$, and the second bound results from our specific choice for the discrete Hubbard-Stratonovich transformation (in both cases an extra factor of $1/10$ was included based on our comparison with an exact calculation, as described in Section~\ref{TestsCrossChecks}.)
The Monte Carlo autocorrelation length was estimated (by computing the autocorrelation function of the total energy) to be approximately 200 Metropolis steps at $T\approx 0.2 \varepsilon_F$. Therefore, the statistical errors are of 
the order of the size of the symbols in the figure. The chemical potential was chosen so as to have a total of about 45 particles for the $8^3$ lattice. We have performed however calculations for particle numbers ranging from 30 to 80, for lattice sizes $8^3$ and $10^3$ and various temperatures, and in all cases the results agreed within discussed above errors and systematics.
In all runs the single-particle occupation probabilities of the highest energy states were well below a percent for all temperatures. This can be seen in Fig.~\ref{occup}, where the dispersion in the data at fixed momentum is the result of statistical errors and the fact that the lattice is not spherically symmetric.

We tried to extract from our data the asymptotic behavior in the limit of large momenta $n(k)\propto C (k_F/k)^n$, which according to the theory \cite{shina,braaten} should at all temperatures be governed by the same exponent, namely $n=4$. Our results are consistent with a value of  the exponent $n=4.5(5)$, thus in reasonable agreement with the theoretical expectation.

All the quantities computed present a number of common features that are easily 
identified, in particular a low and a high temperature regime, separated by a characteristic temperature 
that we estimate to be $T_0 = 0.23(2)\varepsilon_F$. We shall discuss in the next sections whether $T_0$ can be interpreted as the critical temperature $T_c$ for the onset of superfluidity.

\subsection{Energy and chemical potential}

\begin{figure}[htb]
\centering
\includegraphics[width=7.5cm]{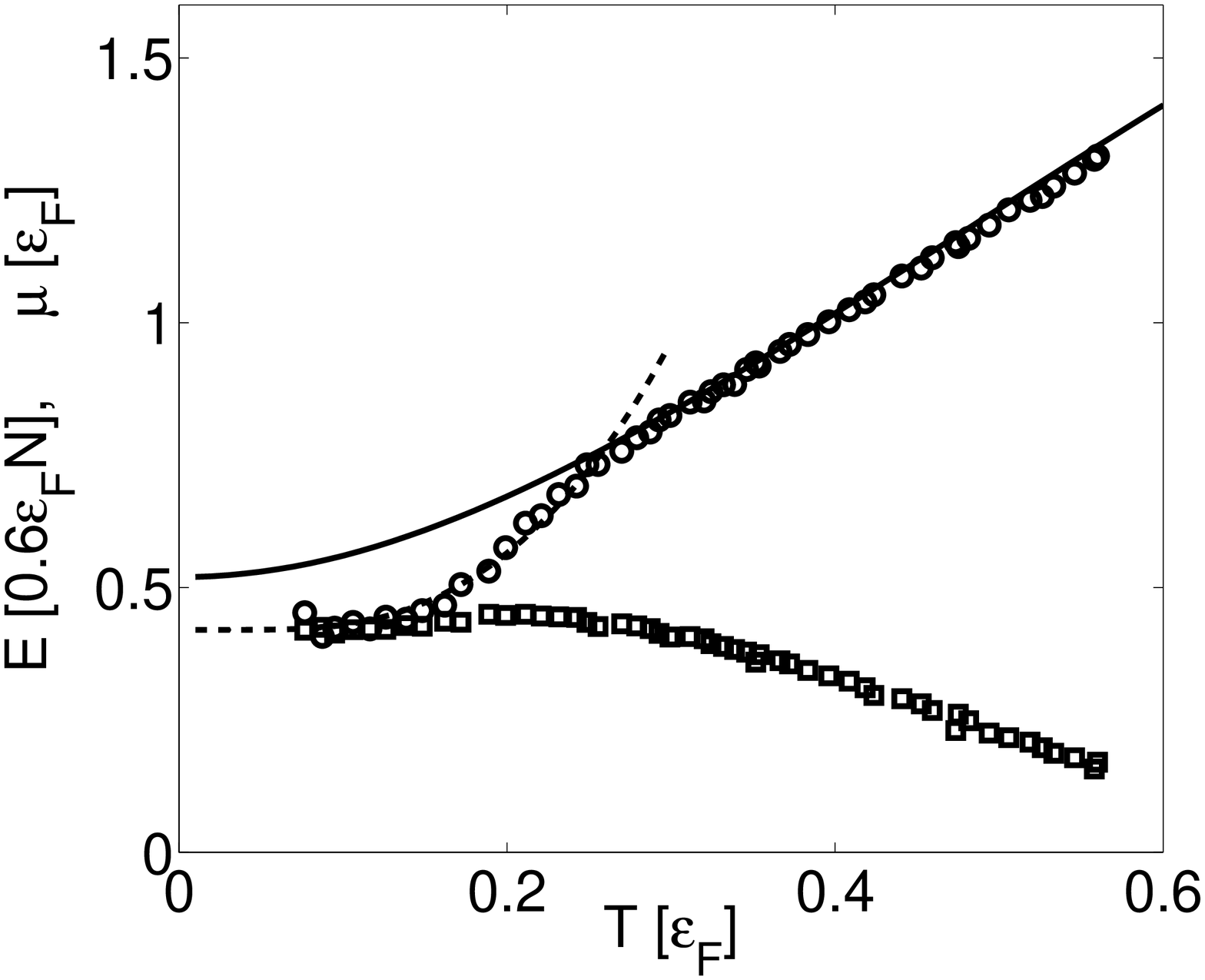}
\includegraphics[width=8.5cm]{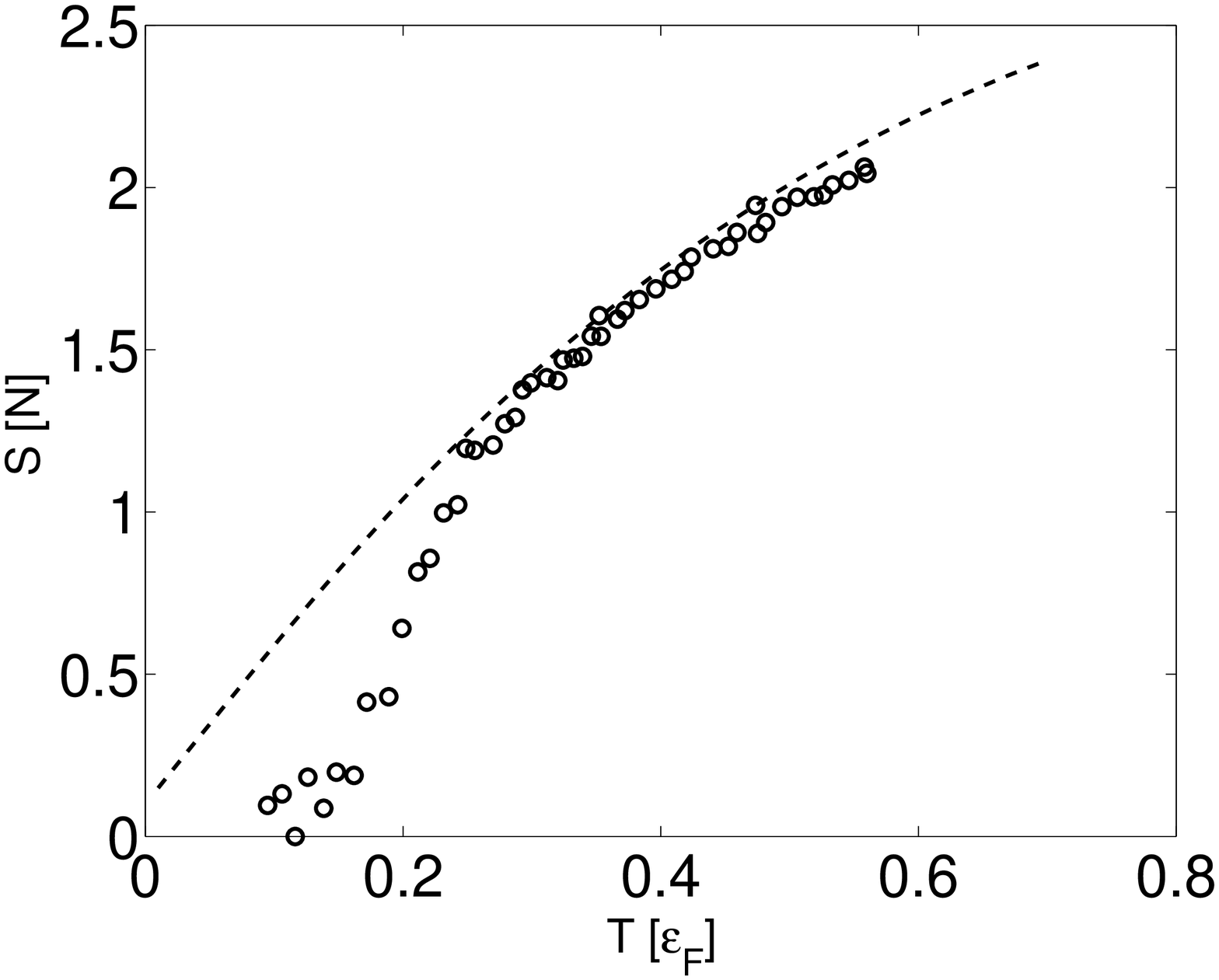}
\caption{ \label{energyentropy} Left panel: total energy $E(T)$ with open circles, and the chemical potential $\mu(T)$ with squares, both for the case of an $8^3$ lattice. The combined Bogoliubov-Anderson phonon and fermion quasiparticle contributions $E_{ph+qp}(T)$ (Eq.~(\ref{eq:phqp})) is shown as a dashed line. The solid line represents the energy of a free Fermi gas, with an offset (see text).
Right panel: entropy per particle with circles for $8^3$ lattice, and with a dashed line the entropy of a free Fermi gas with a slight vertical offset. The statistical errors are the size of the symbol or smaller.}
\end{figure}

As $T \rightarrow 0$ the energy tends to the $T=0$ results obtained by other groups \cite{carlson,chang,giorgini}. This confirms such results, as the algorithms they used (namely Green Function Monte Carlo) are constrained by the existence of a sign problem, which is not the case in the present approach. The solid line in Fig.~\ref{energyentropy} shows the energy of a free Fermi gas, with a shift down given by $1-\xi_n$, with $\xi_n = \xi_s + \delta \xi \simeq 0.55$, where $\xi_s = 0.4$ and
\begin{equation}
\delta \xi =
\frac{\delta E}{\frac{3}{5} \varepsilon_F N} = 
 \frac{5}{8} \left (\frac{\Delta}{\varepsilon_F}\right)^2 \simeq 0.15
\end{equation}
is the condensation energy in units of the free gas ground state energy. Here we have used the BCS result $\delta E = \frac{3}{8}\frac{\Delta^2}{\varepsilon_F} N$ (see Ref.~\cite{BulgacYu}), where $\Delta$ is the pairing gap found in Ref.~\cite{CarlsonReddy} to be $\Delta \simeq 0.50 \varepsilon_F$. One can also find the value of $\xi_n$ from our data by determining what shift is necessary to make the solid curve in Fig.~\ref{energyentropy} (which corresponds to the free gas) coincide with the high temperature data (where the gas is expected to become normal). We find that such procedure gives $\xi_n \simeq 0.52$, which is roughly consistent with the value quoted above. This number should also be compared with the results of Refs.
\cite{carlson,CarlsonMorales}, namely $\xi_n \simeq 0.54$, and Ref.\cite{Lobo}, that finds $\xi_n \simeq 0.56$.

For $T<T_0$ we observe that the temperature dependence of the energy can be accounted for by the elementary excitations present in the system in the superfluid phase: boson-like Bogoliubov-Anderson phonons and fermion-like gapped Bogoliubov quasiparticles. Their contribution is given by

\begin{eqnarray}\label{ph+qp}
E_{ph+qp}(T) &=& \frac{3}{5}\varepsilon_FN \left [ \xi_s + \frac{\sqrt{3}\pi^4}{16\xi_s^{3/2}}
\left(\frac{T}{\varepsilon_F}\right )^4  + \frac{5}{2}\sqrt{\frac{2\pi\Delta^3T}{\varepsilon_F^4}}
\exp\left (-\frac{\Delta}{T}\right)\right ] ,\label{eq:phqp} \\
\Delta &\approx& \left (\frac{2}{e}\right )^{7/3} \!\!\!\!\!\! 
\varepsilon_F \exp \left (\frac{\pi}{2k_Fa}\right ),
\end{eqnarray}
where $\Delta$ is the approximate value of pairing gap at $T=0$ determined in Ref. \cite{carlson} to be very close to the weak-coupling prediction of Gorkov and Melik-Barkhudarov \cite{gorkov}, and $\xi_s\approx 0.44$ \cite{chang} and $\varepsilon_F=\hbar^2k_F^2/2m$ and $n= k_F^3/3\pi^2$ respectively. Notice that the estimate for central value of $\xi_s$ has decreased in the last years: Ref.~\cite{CarlsonReddy} reported $\xi_s = 0.42(2)$, while recently $\xi_s= 0.40(1)$ has been reported in Ref.~\cite{GezerlisCarlson,zhang}, and $\xi_s=0.37(5)$ in this work (see Table \ref{table:results} below), even though all these results agree within quoted errors. The sum of the contributions from the excitations, namely Eq.~\ref{ph+qp}, is plotted in Fig.~\ref{energyentropy} as a dashed line. Both of these contributions are comparable in magnitude over most of the temperature interval $(T_0/2,T_0)$. Since the above expressions are only approximate formulas for $T\ll T_c$, the agreement with our numerical results may be coincidental.

At $T>T_c$ the system is expected to become normal. If $T_0$ and $T_c$ are identified, then the fact that the specific heat is essentially that of a normal Fermi liquid $E_F(T)$ above $T_0$ is somewhat of a surprise, as one would expect the presence of a large fraction of non-condensed unbroken pairs. Indeed, the pair-breaking temperature has been estimated to be $T^* \simeq 0.55\varepsilon_F$, based on fluctuations around the mean-field, see Refs.~\cite{pieri,leggett}, implying that for $T_c < T < T^*$ there should be a noticeable fraction of non-condensed pairs. As we shall see, $T_c \simeq 0.15 \varepsilon_F < T_0$ (this result for $T_c$ was first obtained in Ref.\cite{burovski}), and as one can see in the caloric curve of Fig.~\ref{energyentropy}, the specific heat deviates from the normal Fermi gas, in the range $T_c < T < T_0$.

On the other hand, the chemical potential $\mu$ is almost constant for $T<T_0$, a fact reminiscent of the behavior of an ideal Bose gas in the condensed phase, even though in such a phase our system is strongly interacting and superfluid. This implies a strong suppression of fermionic degrees of freedom at those temperatures. Moreover, assuming $\mu(T) = const.$ for $T<T_0$ implies that
\begin{equation}
E(T) = N \frac{3}{5}\varepsilon_F\xi\left(\frac{T}{\varepsilon_F}\right), \ \ \ \ \  \xi\left(\frac{T}{\varepsilon_F}\right) 
= \xi_s + \zeta \left(\frac{T}{\varepsilon_F}\right)^n, \ \ \ \ \ n = \frac{5}{2}
\end{equation}
which is the temperature dependence of an ideal Bose condensed gas. According to our QMC results, the value of $n$ extracted from our data is $n = 2.50(25)$. 

\subsection{Entropy}

From the data for the energy $E$ and chemical potential $\mu$ one can compute the entropy $S$, because in the unitary limit the relation $PV = \frac{2}{3}E$ holds (true of a free gas as well), where $P$ is the pressure, $V$ is the volume and $E$ is the energy. It is straightforward to show that
\begin{figure}[htb]
\includegraphics[scale=0.49]{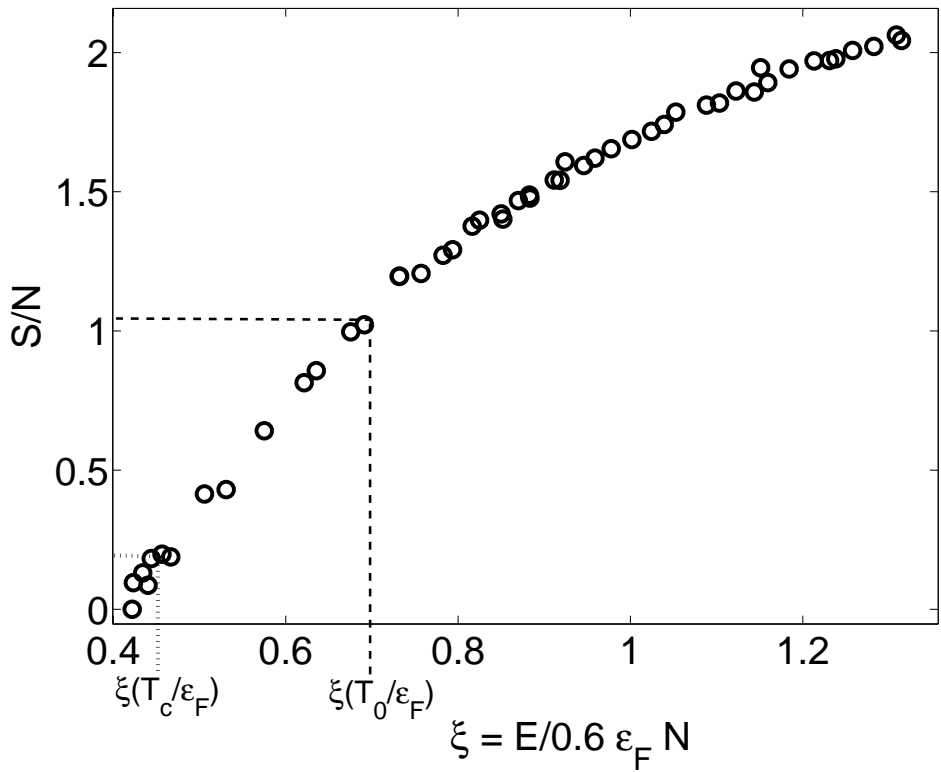}
\caption{\label{SvsE} Entropy per particle as a function of the energy per particle $\xi$. The dotted line shows the location of $T_c$ (see next two sections); the dashed line shows the location of $T_0$.}
\end{figure}

\begin{equation}
\frac{S}{N} = \frac{E + PV - \mu N}{N T} = \frac{\xi(x) - \zeta(x)}{x}
\end{equation}
where $\zeta(x) = \mu/\varepsilon_F$ and $x = T/\varepsilon_F$
which determines the entropy per particle in terms of quantities we know from our simulation. The entropy also departs from the free gas behavior as the temperature is lowered below $T_0$. 

As indicated in \cite{usII}, this data can be used to calibrate the temperature scale at unitarity. Indeed, extending the suggestion of Ref.~\cite{carr}, from known $T$ in the BCS limit, the corresponding $S(T_{BCS})$ can be determined. Then, 
by adiabatically tuning the system to the unitary regime, one uses $S(T_{BCS}) = S(T_{unitary})$ to determine $T$ at unitarity (In practice the experimental procedure goes in the opposite direction, namely measurements are performed at unitarity and then the system is tuned to the deep BCS side, see Ref.~\cite{Luo}).

On the other hand, knowledge of the chemical potential as a function of temperature (see previous section) allows for the construction of density profiles by using of the local density approximation. In turn, this makes it possible to determine $S(E)$ for the system in a trap, fixing the temperature scale via $\partial S/\partial E = 1/T$. Direct comparison with experimental results, in remarkable agreement with our data, has been demonstrated by us in Ref.~\cite{TrappedUFG}. In Fig.~\ref{SvsE} we show the entropy per particle $S/N$ as a function of the energy per particle $\xi$ for the homogeneous system (See Ref.~\cite{TrappedUFG} for the corresponding result for the case of the unitary gas in a trap.)

\subsection{Two-body density matrix and condensate fraction}

\begin{figure}[htb]
\centering
\includegraphics[scale=0.33]{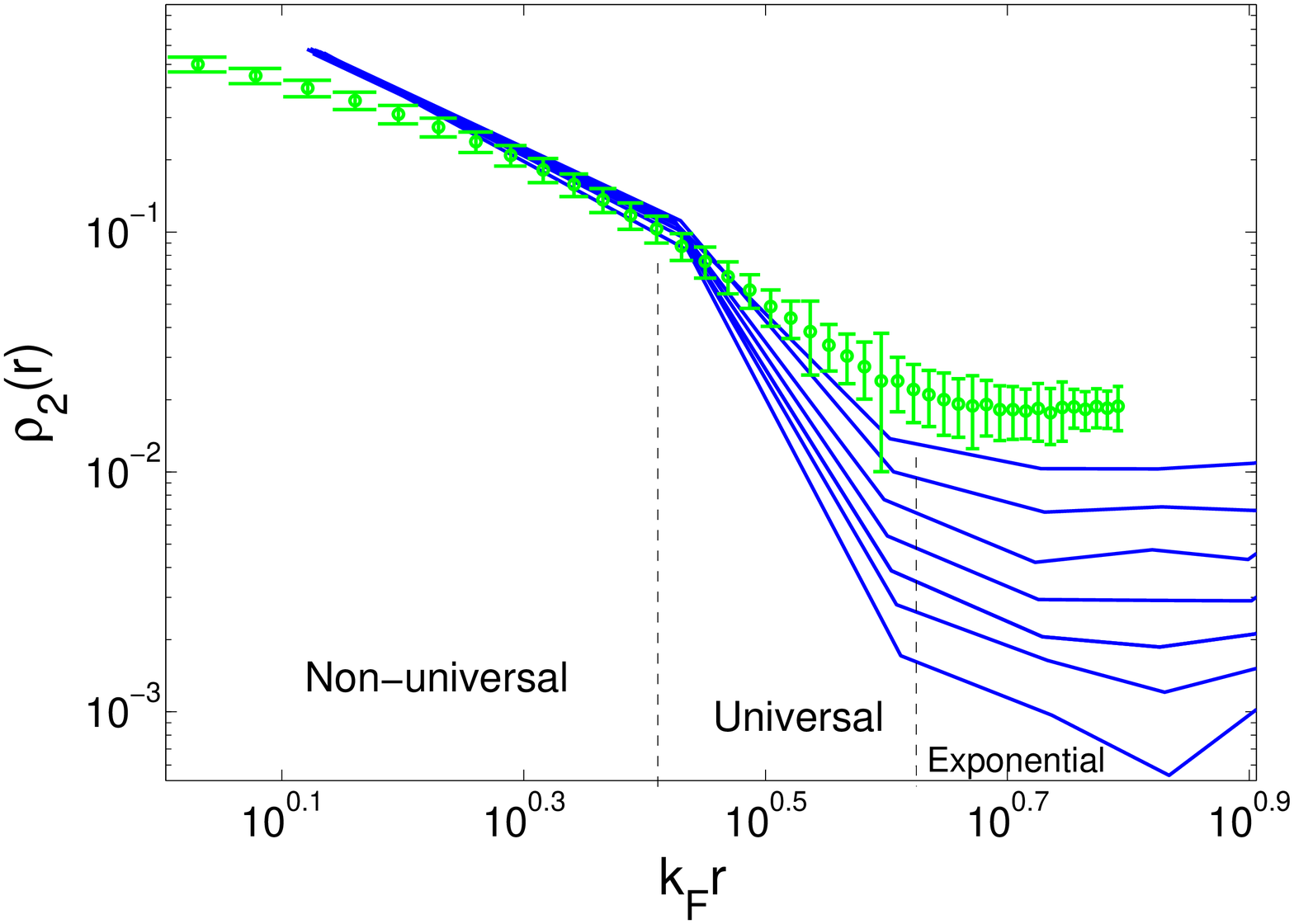}
\includegraphics[scale=0.35]{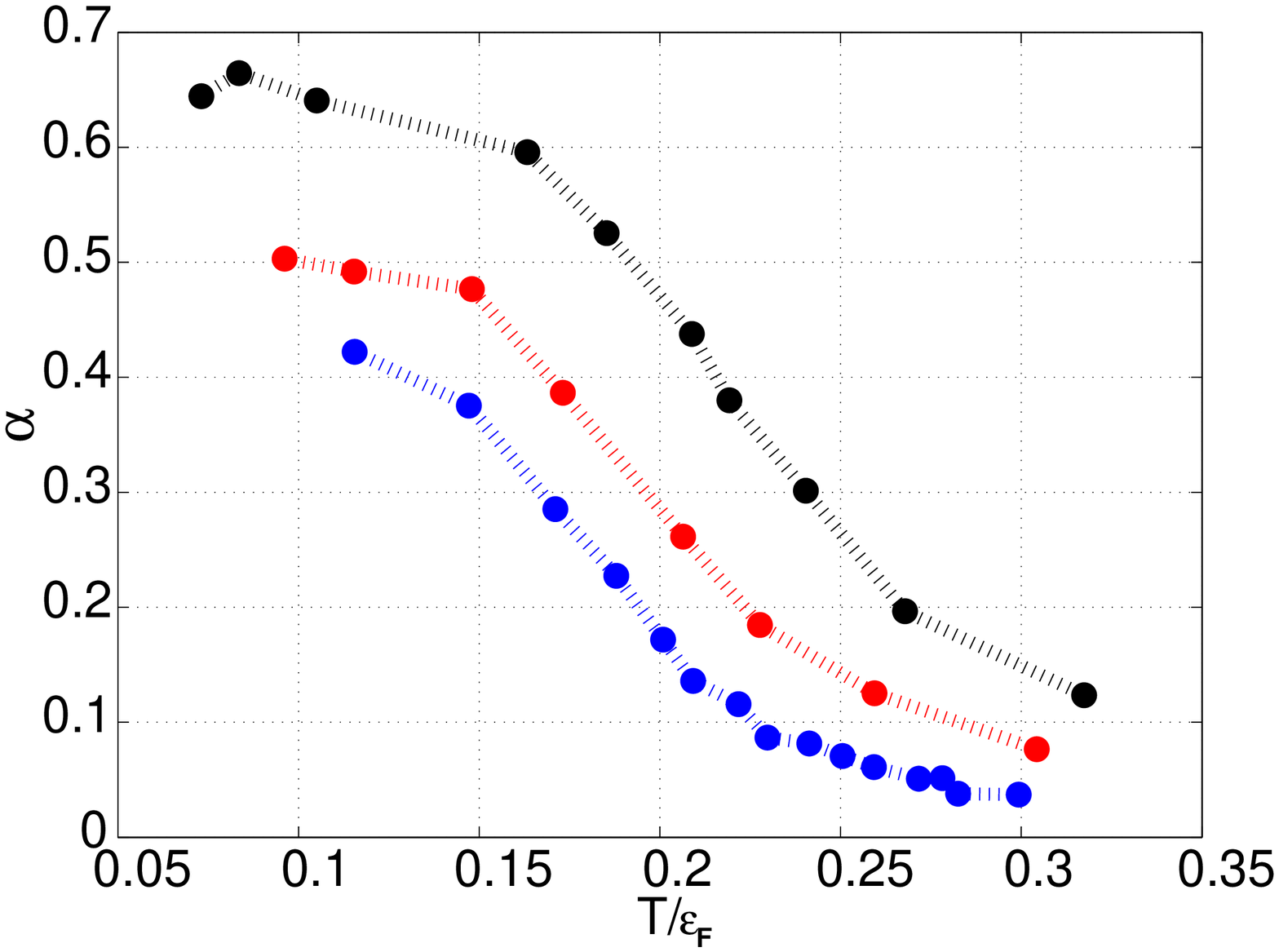}
\caption{\label{CondFrac} (Color online)  Left Panel: Projected two-body density matrix (see text) as a function of position on the lattice. 
In solid blue for a $10^3$ lattice, for the temperature range $(0.1\varepsilon_F,0.3\varepsilon_F)$. The $T=0$ results from Ref. \cite{stefano} are shown with green circles and error bars. Right Panel: Condensate fraction $\alpha(T)$, black for $6^3$ (highest), red  for $8^3$ (middle) and blue for $10^3$ (lowest) lattices, respectively.  }
\end{figure}

Information about the onset of critical behavior (e.g. a superfluid phase transition) can be obtained by studying an appropriate order parameter, both as a function of temperature and system size. In the case of superfluidity in two-component Fermi systems, which is a particular example of off-diagonal long-range order \cite{yang}, the order parameter is the long-distance behavior of the two-body density matrix $g_2(r)$, defined in Eq.~(\ref{TBDM}). At unitarity, knowledge of $g_2(r)$   
is enough to determine the condensate fraction $\alpha = \lim_{r\rightarrow\infty} \frac{N}{2}g_2(r)$.
On the BCS side of the resonance, as discussed in \cite{stefano}, the calculation of $\alpha$ demands also knowledge of the one-body density matrix:
\begin{equation}
\rho(r)=\frac{2}{N}\int d^3{\bf r}_1 
\langle \psi_{\uparrow}^\dagger({\bf r}_1+{\bf r}) \psi_{\uparrow}({\bf r}_1)\rangle 
\label{OBDM}
\end{equation}  
and $\alpha = \lim_{r\rightarrow\infty}\frac{N}{2}g_2(r) - \rho^2(r)$.

\begin{figure}[htb]
\includegraphics[width=12cm]{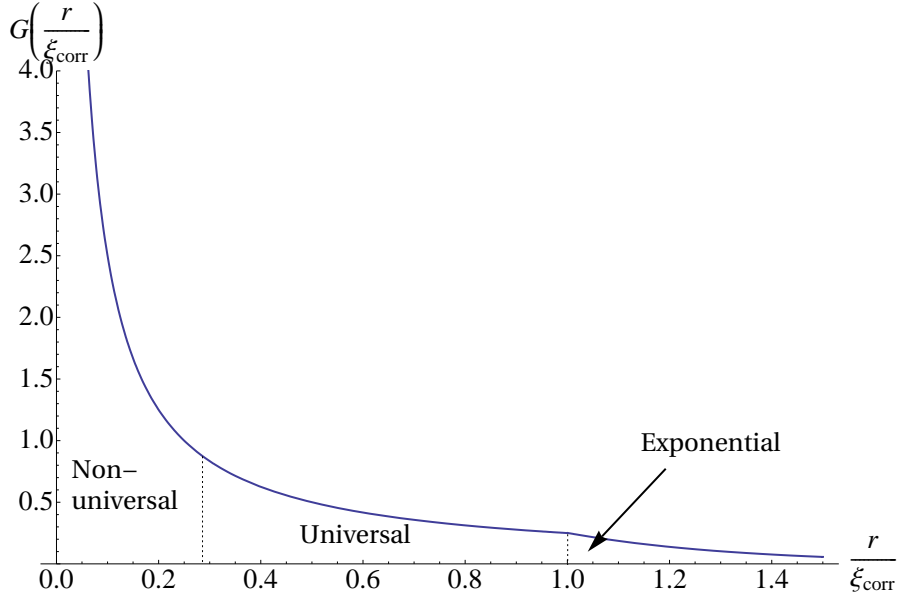}
\caption{\label{CorrFunc} (Color online) Generic form of a correlation function $G(r/\xi_\text{corr})$ as a function of the radial coordinate $r$ in units of the correlation length $\xi_\text{corr}$.}
\end{figure}

In Fig. \ref{CondFrac} we show our results for $g_2(r)$, as a function of the dimensionless lattice position $k_F r$ (left panel) and the extracted condensate fraction $\alpha$ at unitarity for several lattice sizes (right panel). In Fig. \ref{CorrFunc}, on the other hand, we show schematically the generic form of a correlation function $G(r/\xi_\text{corr})$(such as $g_2(r)$). At short distances, $r \sim l$ (where here $l$ should be regarded as an intrinsic short distance scale of the problem), the behavior of $G$ depends on the system under consideration, i.e., it is non-universal. In the region $l \ll r < \xi_\text{corr}$ (which extends to infinity at a phase transition because there $\xi_\text{corr} \rightarrow \infty$, see next section) the form of $G$ is universal, in the sense that it depends, quite generally, only on the spatial dimensionality of the problem and the internal symmetries of the Hamiltonian. In that region $G \sim r^{-(1+\eta)}$, where $\eta \simeq 0.038$ is a universal critical critical exponent (see e.g.~\cite{LandauLifshitzStatMech}).  Finally, for $r > \xi_\text{corr}$ the correlation decays exponentially with $r/\xi_\text{corr}$.

Below the critical temperature, where $\alpha$ is non-zero, its value can be extracted from the asymptotic form of $g_2(r)$. In Fig.~\ref{CondFrac} (left panel), one can see that the asymptotic value of $g_2(r)$ for $10^3$ is approached even for $r < L/2$, so the lattice sizes used are reasonable large to determine $\alpha$. It is possible that for larger lattice sizes $\alpha$ could become smaller. On the other hand, there is virtually no room for finding the 
power law that characterizes the universal critical behavior of this function (see Fig. \ref{CorrFunc}). In other words, corrections to universal scaling will be important close to the critical point for these small volumes (see next section).

\subsection{Finite size scaling and the critical temperature
\label{sec:FSSandTc}}

By definition, the correlation length characterizing the non-local degree of correlation of a system diverges at a critical point. Moreover, close enough to the transition it diverges as
\begin{equation}
\xi_\text{corr} \propto |t|^{-\nu}
\end{equation}
where $t =1 - T/T_c$  and $\nu$ is a universal critical exponent. For the $U(1)$ universality class (which contains superfluid phase transitions), this exponent is well-known: $\nu = 0.671$.

When dealing with systems that have a finite size $L^3$, the theory of the 
renormalization group (RG) predicts a very specific behavior for the correlation functions
close enough to the transition temperature (see e.g. Ref.~\cite{DombLebowitz8}). In particular, the two-body density 
matrix $K(L,T)$ that gives the order parameter for off-diagonal long-range order, scales as
\begin{equation}
R(L,T) = L^{1+\eta}K(L,T) = f(x)(1+c L^{-\omega}+...)
\end{equation}
where $\eta=0.038$ is another universal critical exponent, $f(x)$ is a universal analytic function, 
$x = (L/\xi_{corr})^{1/\nu}$, and $c$ is a non-universal constant, and $\omega \simeq 0.8$ is the critical 
exponent of the leading irrelevant field. One should keep in mind that typically one knows neither $c$ nor $T_c$, but is interested in finding the latter.

In a typical Monte Carlo calculation $K(L,T)$ is computed for various lengths $L_i$ and
temperatures $T$. The procedure to locate the critical point (characterized by scale invariance) involves finding
the ``crossing'' temperatures $T_{ij}$, for which $R(L_i,T_{ij}) = R(L_j,T_{ij})$ at two given lengths $L_i$ and $L_j$. 
Assuming that one is close to the transition (so that the correlation length is large compared to any other scale), 
one can expand $f(x(|t|)) = f(0) + f'(0)L^{1/\nu}b|t|$ (where $\xi_{corr} = b |t|^{-\nu}$ was used),
and derive the relation
\begin{equation}
|T_c -T_{ij}| = \kappa g(L_i,L_j)
\end{equation}
where
\begin{equation}
g(L_i,L_j) = L_j^{-(\omega + 1/\nu)}
\left[
\frac{\left(\frac{L_j}{L_i}\right)^\omega -1}{1-\left( \frac{L_i}{L_j}\right)^{1/\nu}}
\right]
\end{equation}
and $\kappa = c T_c f(0)/bf'(0)$. If there were no non-universal corrections to scaling (i.e. if $c=0$), then $\kappa = 0$ and $T_c = T_{ij}$, which means that, upon scaling by the appropriate factor (as above) all the curves $K(L,T)$ corresponding to different $L$'s would cross exactly at $T_c$. In general these corrections are present, and it is therefore necessary to perform a linear fit of $T_{ij}$ vs. $g(L_i,L_j)$ and extrapolate to infinite $L$ in order to determine the true $T_c$. Following such procedure our data for the condensate fraction of the unitary Fermi gas indicates that $T_c \lesssim 0.15(1) \varepsilon_F$, considerably lower than the characteristic temperature $T_0 = 0.23(2)$ found by studying the behavior of the energy and the chemical potential.
Even though this result for $T_c$ is close to estimates by other groups (see e.g. \cite{burovski}), it should be pointed out that the experimental data of Ref.~\cite{Luo} shows a distinctive feature in the energy versus entropy curve at a temperature close to $T_0$ (see Ref.~\cite{TrappedUFG}), whereas a clear signature of a transition at a lower temperature remains to be found.
\begin{figure}[htb]
\centering
\includegraphics[scale=0.40]{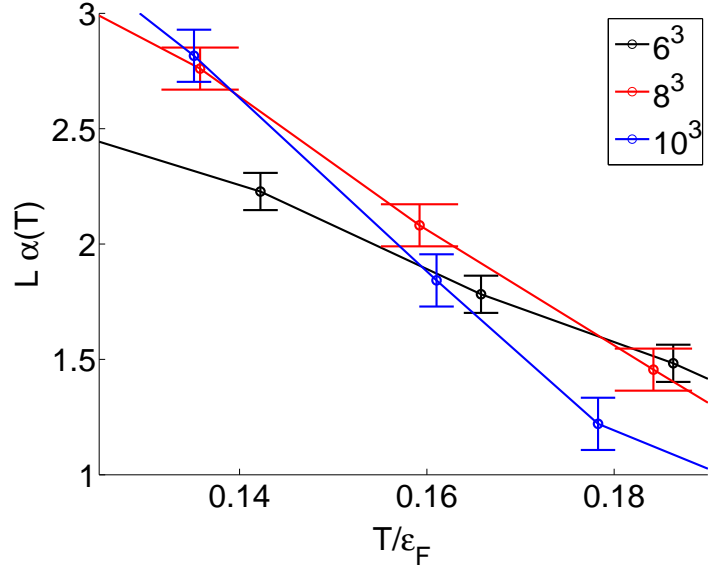}
\caption{\label{Crossings} (Color online)  Condensate fraction $\alpha(T)$ scaled with the lattice size $L$, black squares for $6^3$, red stars for $8^3$ and blue circles for $10^3$ (highest on the left) lattices, respectively for the case $1/k_Fa=-0.1$. The errorbars correspond to the statistical errors.}
\end{figure}

\section{Results away from unitarity}

In the following we describe the results of our calculations away from unitarity. The system was placed on a lattice of volume $V=(8 l)^3$, filled with $N=45\pm 15$ particles.  In all cases the temperatures cover the range $0.07 \leq T/\varepsilon_F \leq 0.5$, corresponding to $N_{\tau}$ steps in the imaginary time direction varying from $N_{\tau}\simeq1700$ to $N_{\tau}\simeq200$, respectively. The temperature is limited from below by the precision of our computers (because the matrices involved become ill-conditioned in the sense explained in section III B), and from above by the fact that our phase space has a natural UV cutoff given by the inverse lattice spacing. In all cases the occupation of the high-energy modes smaller than 1$\%$ percent. The coupling strength was varied in the range $-0.5 \leq 1/k_F a \leq 0.2$ (where $k_F = (3 \pi^2 n)^{1/3}$), and is limited on the negative (BCS) side by the finite volume $V$ (which may become comparable to the size of the Cooper pairs deep in the BCS regime), and on the positive (BEC) side by the finite lattice spacing $l$ (whose size eventually becomes inadequate to describe localized dimers of size $a={\cal{O}}(l)$, deep in the BEC regime, and which manifests itself as poor convergence of observables). The number of uncorrelated Monte Carlo samples varies from $7500$ at the lowest temperatures to $2500$ at the highest. The Monte Carlo auto-correlation time was $\simeq 200$ samples (estimated by studying the autocorrelation of the energy), implying a statistical error of less than $2\%$.

\begin{table}[ht]
\begin{center}
\begin{tabular}{||c||c||c|c|c||c|c|c||} \hline
$1/k_Fa$   & $E(0)/E_F$    & $T_0$    &$\mu_0/\varepsilon_F$ & $E_0/E_F$ & $T_c < $ & $\mu_c/\varepsilon_F$ & $E_c/E_F$ \\
\hline
\hline
-0.5 & 0.60(4)  &   0.14(1)     & 0.685(5)     & 0.77(2)    & --         & --       & --      \\
-0.4 & 0.59(4)  &   0.15(1)     & 0.65(1)      & 0.75(1)    & --         & --       & --      \\
-0.3 & 0.55(4)  &   0.165(10)   & 0.615(10)    & 0.735(10)  & 0.105(10)  & 0.61(1)  & 0.64(2) \\
-0.2 & 0.51(4)  &   0.19(1)     & 0.565(10)    & 0.725(10)  & 0.125(10)  & 0.56(1)  & 0.61(2) \\
-0.1 & 0.42(4)  &   0.21(2)     & 0.51(1)      & 0.71(2)    & 0.135(10)  & 0.50(1)  & 0.54(2) \\
0    & 0.37(5)  &   0.23(2)     & 0.42(2)      & 0.68(5)    & 0.15(1)    & 0.43(1)  & 0.45(1) \\
0.1  & 0.24(8)  &   0.26(3)     & 0.34(1)      & 0.56(8)    & 0.17(1)    & 0.35(1)  & 0.41(1) \\
0.2  & 0.06(8)  &   0.26(3)     & 0.22(1)      & 0.39(8)    & 0.19(1)    & 0.21(1)  & 0.25(1) \\
\hline
\end{tabular}
\end{center}
\caption{\label{table:results} Results for the ground state energy, the characteristic temperature $T_0$, and the corresponding chemical potential and energy, from the caloric curves of Fig.~\ref{fig:EofT}, and the upper bounds on the critical temperature $T_c$ from finite size scaling and the corresponding chemical potential and energy.}
\end{table}

\begin{figure}[htb]
\includegraphics[scale=0.27]{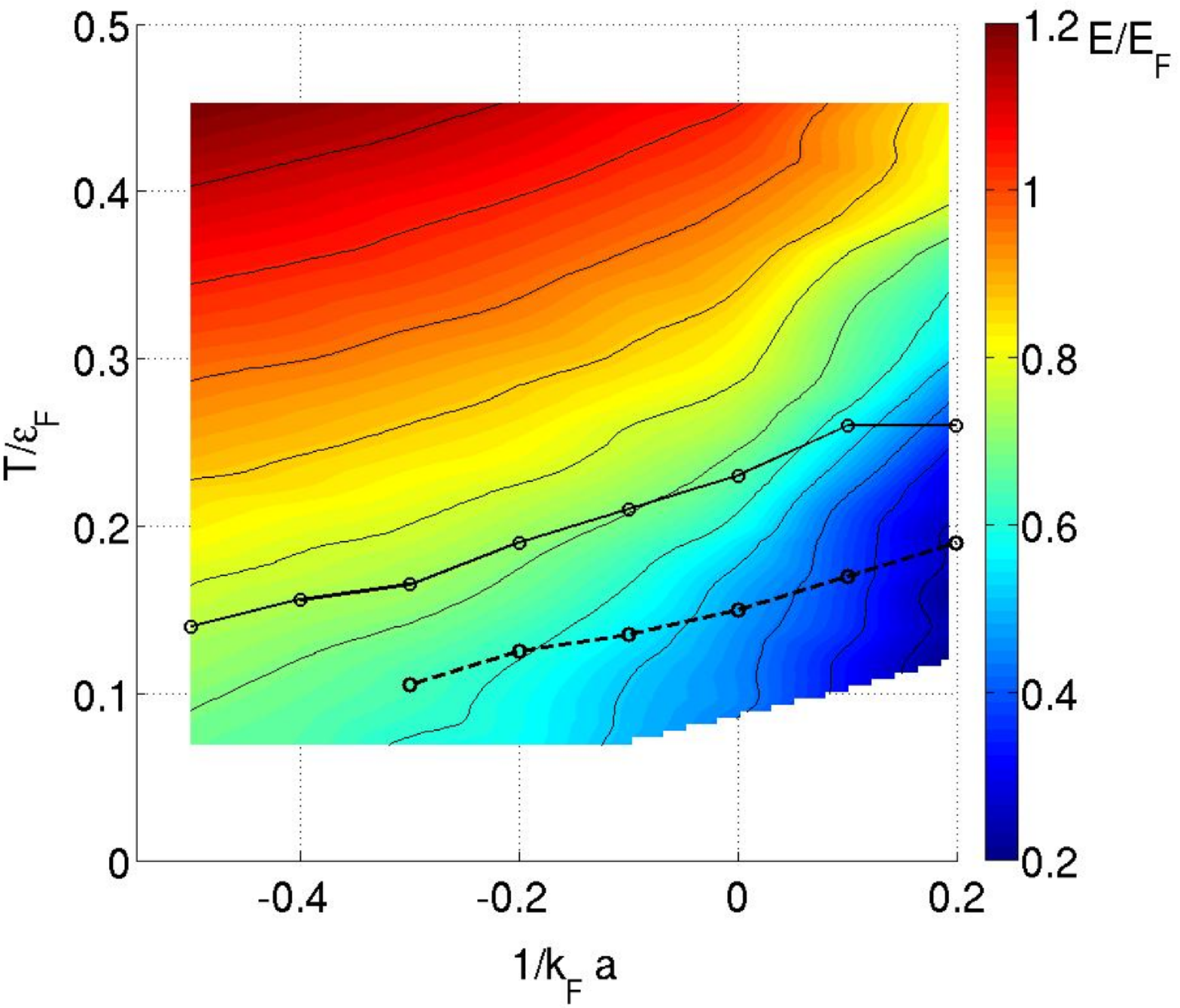}
\includegraphics[scale=0.27]{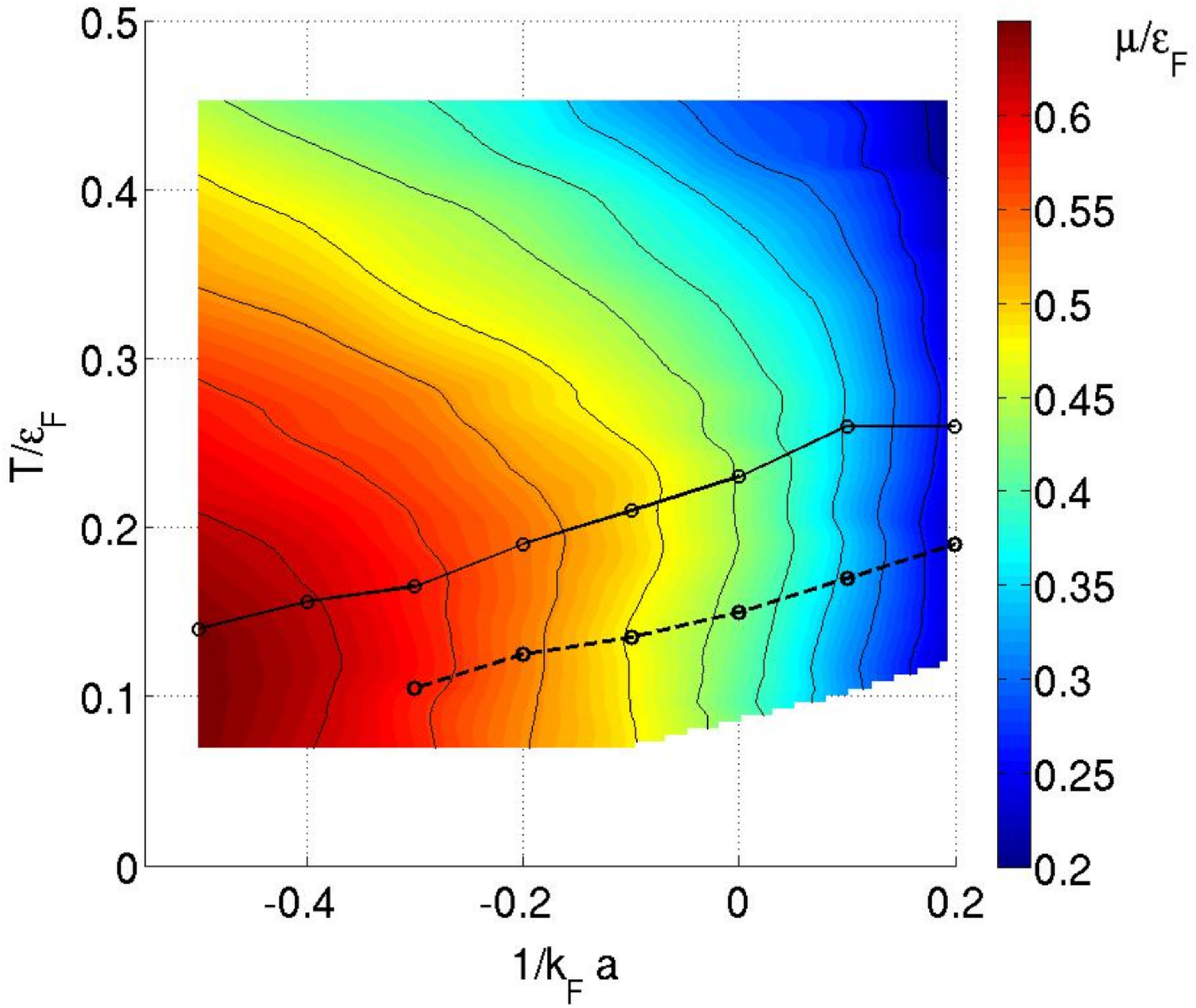}
\caption{\label{fig:EofT} (Color online) Energy (left panel) and chemical potential (right panel) as functions of $T/\varepsilon_F$ and $1/k_Fa$. The dashed line shows the location of $T_c$ and the solid line represents $T_0$, both as functions of $1/k_Fa$, from Table~\ref{table:results}.}
\end{figure}



Repeating the analysis of Sec.\ref{sec:FSSandTc} one arrives at the estimates for $T_c$ shown in Table~\ref{table:results}. Notice however that, given the rather small lattice sizes used (the limitation being given by the required computer power/time, and ultimately by its scaling with the size of the lattice), an extrapolation to $L\rightarrow\infty$ is difficult. Still, the study of the crossing temperatures $T_{ij}$ provides us at least with upper bounds on $T_c$, which is what we show in Table~\ref{table:results}. Unfortunately, it was not possible for us to explore temperatures below $0.1\varepsilon_F$, and therefore we were unable to find $T_c$ on the BCS side beyond $(k_F a)^{-1} = -0.3$.
The fact that our upper bound on $T_c$ at unitarity agrees with extrapolations to $L\rightarrow\infty$ performed by other groups \cite{burovski} indicates that these bounds are not far from the actual result. In Ref. \cite{njp} the authors performed a finite size scale analysis of our initial data \cite{bdm} and found a value of $T_c$ in agreement with their result. In Fig. \ref{Crossings} we show an example of the finite size scaling analysis at $1/k_Fa=-0.1$ performed as explained in Sec.\ref{sec:FSSandTc}. 
The last two columns of Table~\ref{table:results} show the chemical potential and the energy at the value of the bound on the critical temperature. In particular, the values at unitarity, namely $(k_Fa)^{-1} = 0; \mu_c/\varepsilon_F = 0.43(1) ; E_c/E_F = 0.45(1)$ should be compared with the results of reference \cite{burovski}: $\mu_c/\varepsilon_F=0.493(14); E_c/E_F=0.52(1)$, both of which are higher than our estimates. It should also be pointed out that the latter values, shown in Fig.~\ref{hT}, slightly violate the bounds imposed by thermodynamic stability (see Appendix), which is not the case for our data.

\begin{figure}[h]
\includegraphics[scale=0.45]{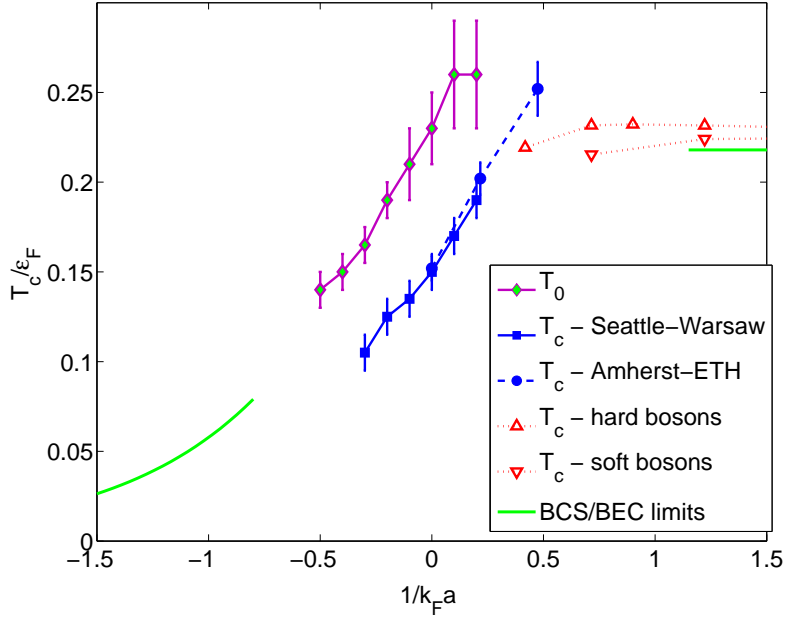}
\caption{\label{Tcnew} (Color online) The solid (purple) curve and diamonds represent the temperature $T_0$ at which the energetic behavior of the system shows a transition from Bose-like to Fermi-like behavior.  
The solid (green) curve in the low-left corner is $T_c$ estimated within the BCS theory with the Gorkov-Melik-Barkhudarov correction \cite{gorkov}. The solid (green) curve on the right is the $T_c\approx 0.218 \varepsilon_F$ for a non-interacting Bose gas. The (red) up-triangles and the down-triangles joined by a dotted line respectively are the $T_c$ for the hard-core and soft-core bosons calculated in Ref. 
\cite{pilati}. The three (blue) dots with error bars joined by a dashed line are the results of Ref. \cite{newdata}, while the six (blue) squares joined by a solid line are our new estimates for $T_c$ presented din this work.   }
\end{figure}

Recently the Amherst-ETH group has posted values $T_c$ for a couple of values of the coupling constant $1/k_Fa\ge 0$ with some of the details of the calculations, see Ref. \cite{newdata} and Fig. \ref{Tcnew}.

The strong dependence observed by these authors earlier on the filling factor, see discussion around Fig. (\ref{DispersionRelation}), was presumably due to the use of an inaccurate representation of the kinetic energy, which becomes accurate only at very low filling factors (when $k_F<1/l$); see, however, our discussion on density dependence in Sec. \ref{sec:densitydependence}.
The values of the critical temperature estimated in this work and in Ref. \cite{newdata} agree within the error bars at unitarity and at $1/k_Fa \approx 0.2$. The value  $T_c/\varepsilon_F=0.252(15)$ at $1/k_Fa=0.474(8)$ \cite{newdata} does not seem to follow the systematics suggested by the rest of the results for 
$1/k_Fa\le 0.22$, the critical temperature for the hard and soft bosons \cite{pilati} and the limiting BCS and BEC behavior.  If one ignores the value $T_c/\varepsilon_F=0.252(15)$ at $1/k_Fa=0.474(8)$ \cite{newdata}, the data presented in Fig. \ref{Tcnew} would thus suggest that for the value of the coupling constant $1/k_Fa\approx 0.8$ the critical temperature attains a maximum of $T_c/\varepsilon_F \approx 0.23(2)$.

The results for the energy $E$ per particle (in units of the free gas ground-state energy $E_F = 3/5 \varepsilon_F N$, where $N$ is the total number of particles) are shown in Fig.~\ref{fig:EofT}, along with the chemical potential $\mu$ (in units of the free gas Fermi energy $\varepsilon_F$). 
For every value of $1/k_Fa$ that we studied, our data presents two salient features: below certain temperature $T_0$, $\mu/\varepsilon_F$ is approximately constant (a feature of free Bose gases in the condensed phase); above that temperature $\mu/\varepsilon_F$ decreases steadily, while $E/E_F$ becomes the energy of a free Fermi gas, offset by a constant energy (whose specific value depends on $1/k_Fa$). In Table \ref{table:results} we summarize our results at $T_0$. The errors represent uncertainties in the point of departure of $E/E_F$ from the (offset) free Fermi gas, and the departure of $\mu/\varepsilon_F$ from its (approximately) constant low-temperature value. In the same table we also show our extrapolated values for the ground state energies. 
At low enough temperature both $E/E_F$ and $\mu/\varepsilon_F$ become approximately constant. From this observation it can be inferred that the thermal fluctuations are small enough that those constant values should not differ greatly from their ground-state values. 
For reference we also include our data for the system at unitarity. The fact that the latter data falls in the right place shows that our calculations are quite close to the dilute limit. In this respect one should also note that the $T=0$ fixed-node Monte-Carlo calculations show a similar agreement, even though in one case $nr_0^3\approx 10^{-3}$  \cite{carlson}, while in the other $nr_0^3\approx 10^{-7}$ \cite{giorgini}, where $r_0$ is the effective range of the interaction used. Notice that our estimated value for $\xi$ at unitarity is lower than the variational estimates reported in Refs. \cite{carlson,chang,giorgini,GezerlisCarlson} and in apparent agreement within error bars with the unpublished results of Ref. \cite{zhang}.


\begin{figure}[htb]
\includegraphics[scale=0.35]{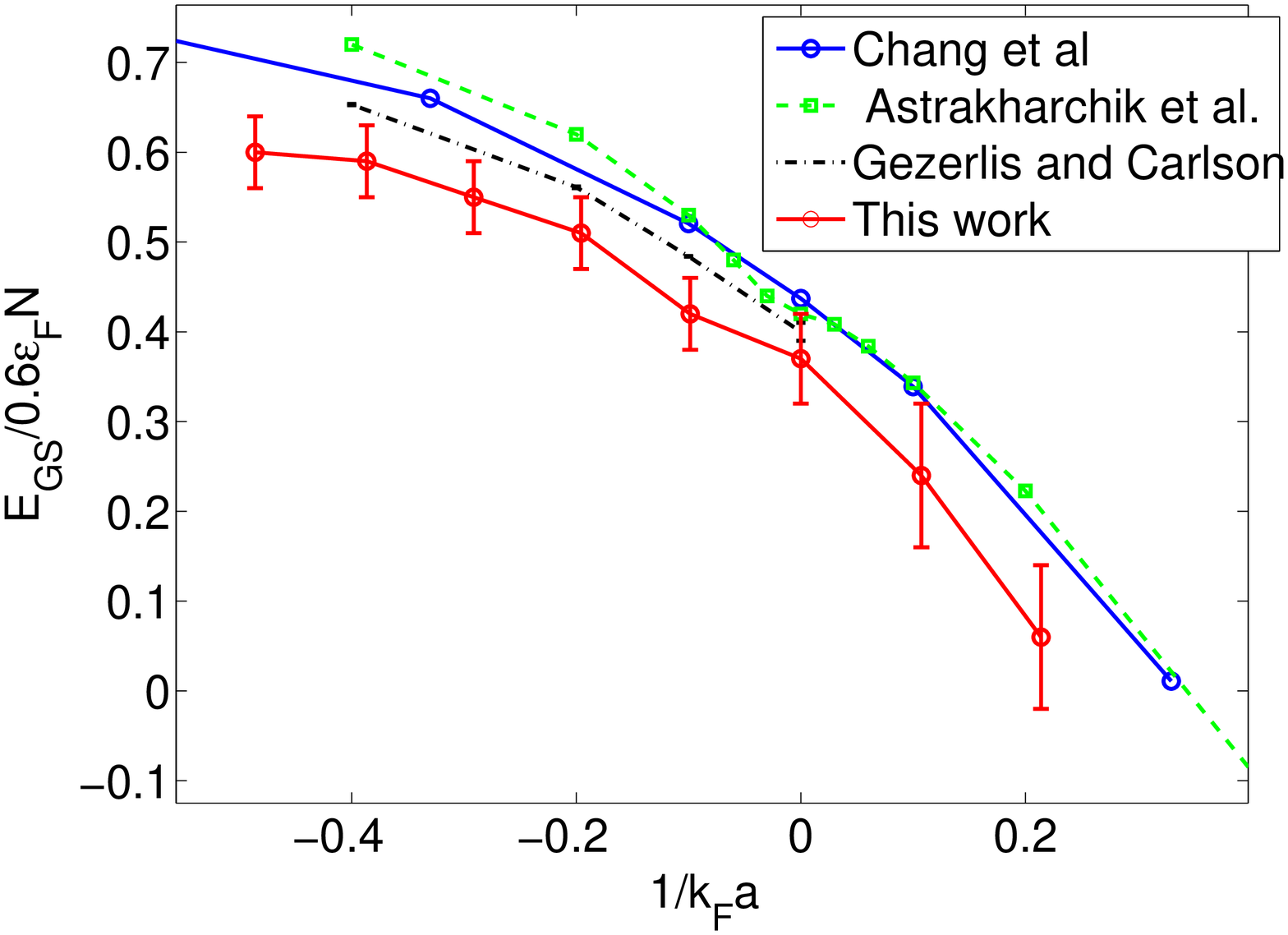}
\includegraphics[scale=0.35]{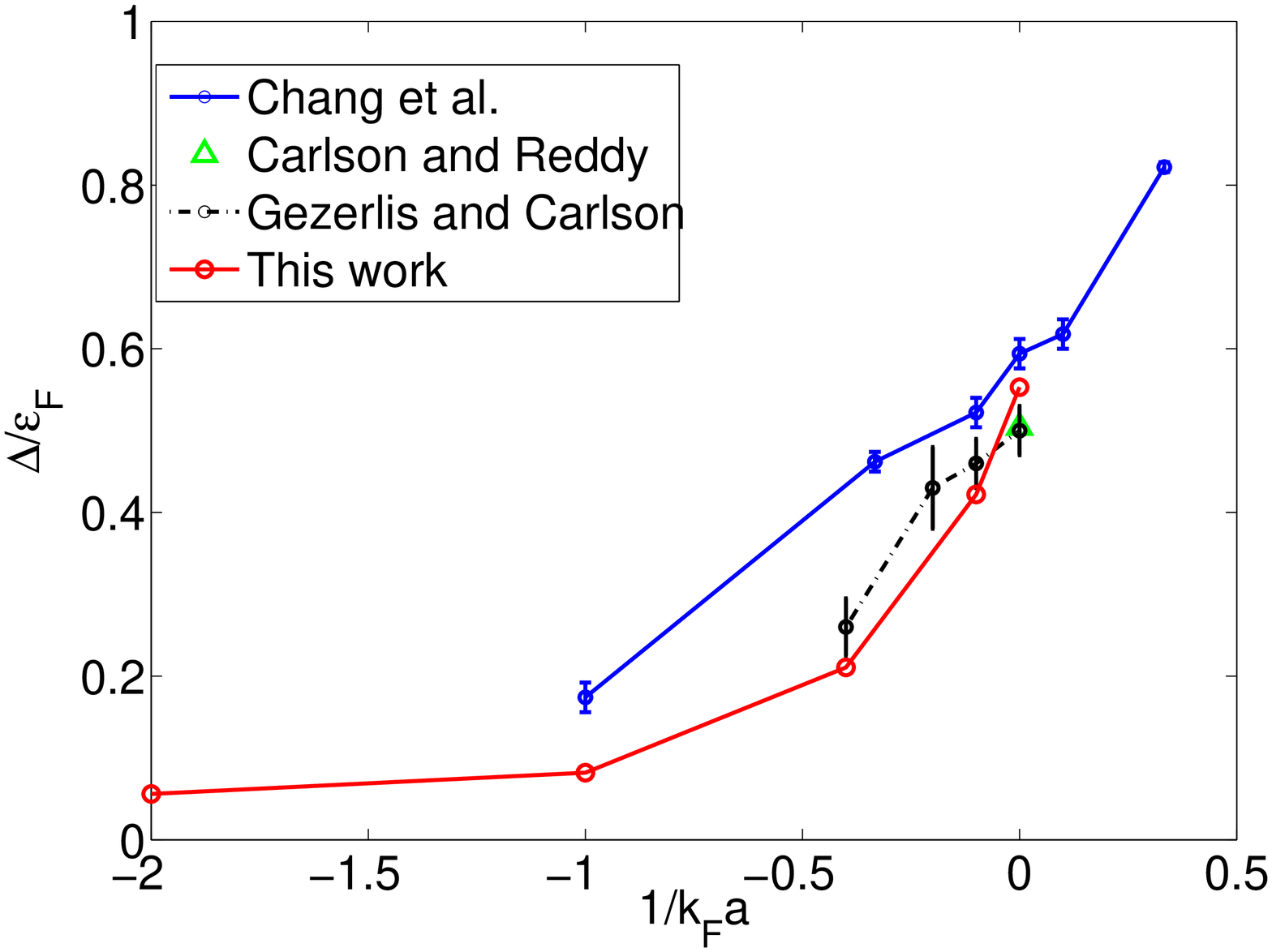}
\caption{\label{GSProperties} (Color online) Ground state energy (left panel, in units of the free gas ground state energy $3/5\varepsilon_F N$), obtained by extrapolating our finite temperature results to $T=0$, is shown in red circles with error bars (representing the uncertainty in the extrapolation). The data of Ref.~\cite{chang} appears in blue circles, Ref.~\cite{GezerlisCarlson} in black dash-dotted line, and Ref.~\cite{giorgini} in green squares. The right panel shows the gap at the lowest temperatures, from our work in Ref.~\cite{SpectrumUFG}, in red circles. Blue circles with error bars show the data of Ref.~\cite{chang}, the black dash-dotted line shows the data of Ref.~\cite{GezerlisCarlson}, and the green triangle represents the data of Ref.~\cite{CarlsonReddy}.}
\end{figure}

In the left panel of Fig.~\ref{GSProperties} we show $E/E_F$ extrapolated to $T=0$, as a function of $1/k_F a$, together with the ground-state energy as determined in Refs.~\cite{chang, stefano, GezerlisCarlson}. 
Since our calculations are in principle exact, it is not suprising that our extrapolations yield results that are consistently lower than those in Refs.~\cite{chang, stefano, GezerlisCarlson}, since those calculations are based on the Fixed-Node approximation, which are thus variational and provide an estimate for the ground state energy from above.

In the right panel of Fig.~\ref{GSProperties} we present the results for the pairing gap at the lowest temperatures. This quantity was determined via a calculation of a response function $\chi$ (see Ref.~\cite{SpectrumUFG}) as a function of momentum $\bf p$:
\begin{equation}
\chi ({\bf p})=
-\int_0^\beta d \tau {\cal{G}}_\beta({\bf p},\tau)
\end{equation}
where 
\begin{equation}
{\cal{G}}_\beta({\bf p},\tau) = 
\frac{{\text{Tr}}[e^{-(\beta-\tau) (H-\mu N)} \psi_{\uparrow}({\bf{p}})  e^{-\tau (H-\mu N)} \psi_{\uparrow}^\dagger({\bf{p}})]}{Z(\beta,\mu,V)} 
\end{equation}
is the temperature Green function. This response function has been shown by us in Ref.~\cite{SpectrumUFG} to be accurately parametrized by the independent quasi-particle form given by
\begin{equation}
\chi ({\bf p})=\frac{1}{E({\bf p})}\frac{e^{\beta E({\bf p})}-1}{e^{\beta E({\bf p})}+1},
\end{equation}
with
\begin{equation}
E({\bf p}) = \sqrt{ \left ( \frac{\alpha p^2}{2m}+U-\mu \right )^2+\Delta^2}.\label{eq:ep}
\end{equation}
In this expression, $\alpha = m/m_*$, where $m_*$ is an effective mass, $U$ is the mean-field potential, $\Delta$ represents the pairing gap, and $\mu$ is the chemical potential. All of these quantities are functions of temperature (see Ref. \cite{SpectrumUFG} for further details). The data for $\Delta$ shown in the right panel of Fig.~\ref{GSProperties} corresponds to the lowest temperatures we have simulated (namely $T\lesssim 0.1\varepsilon_F$). Our results for $\Delta$ agree qualitatively with the data by other groups (\cite{chang,CarlsonReddy,GezerlisCarlson}).  Away from unitarity, however, our data falls systematically below the data by other groups.

\section{Summary and conclusions}

In this paper we have described the technical details involved in the non-perturbative calculation of thermal averages of systems of interacting fermions at finite temperature.
We have performed calculations of the thermal properties of a system of spin 1/2 fermions at and away from the unitary point. The particles were placed on a 3D spatial lattice, in a path integral formulation of the interacting many-body problem. 

By studying the finite size scaling of the condensate fraction we have established upper bounds on the critical temperature $T_c$ of the superfluid-normal phase transition, for couplings around the unitary point in the region $-0.5 \le (k_F a)^{-1} \le 0.2$. At unitarity we find $T_c \simeq 0.15(1)$, which is in agreement with Ref.~\cite{burovski}. In contrast, at the transition the energy $E_c$ and the chemical potential $\mu_c$ that we find are lower than those of Ref.~\cite{burovski} by about $15\%$. Furthermore, we find that $E_c$ and $\mu_c$ of Ref.~\cite{burovski} slightly violate the bounds imposed by thermodynamic stability, which are satisfied by our data, as shown in Fig.~\ref{hT}.

For all the couplings we studied, in particular at unitarity, our results for the universal function $\xi$ and the chemical potential are consistent with normal Fermi gas behavior above a characteristic temperature $T_0 > T_c$ that depends on the coupling. $T_0$ is obtained by studying the deviations of the caloric curve from that of a free Fermi gas. Furthermore, the chemical potential is approximately constant below $T_0$. The existence of such a characteristic temperature that is different from the critical temperature is analogous to the case of water, where density reaches a maximum at a temperature $T \simeq 4 ^{\circ}{\rm C}$, which is above the $T = 0 ^{\circ}{\rm C}$ liquid-solid phase transition.

At unitarity we find $T_0=0.23(2)$, which is in agreement with the experimental results of Ref.~\cite{Luo}, where measurements of the caloric curve and energy vs. entropy curve of a unitary Fermi gas were reported. For $T_c < T < T_0$ there is a noticeable departure from normal Fermi gas behavior, possibly due to pairing effects.

Extrapolations of our data for the energy to $T=0$ are systematically below the results by other groups. This is not surprising because Green Function Monte Carlo methods provide an upper bound to the energy. 

We also compare our low temperature results for the gap (determined through the calculation of a response function, as explained in Ref.\cite{SpectrumUFG}) with ground state calculations and find reasonably good agreement close to the unitary point, and somewhat lower values on the BCS side of the resonance.

\section{Appendix: Thermodynamic relations at unitarity.}

In this section we complete our discussion of thermodynamics at unitarity by deriving a number of identities and expressing the various thermodynamic functions in useful forms.

\begin{figure}[htb]
\includegraphics[scale=0.5]{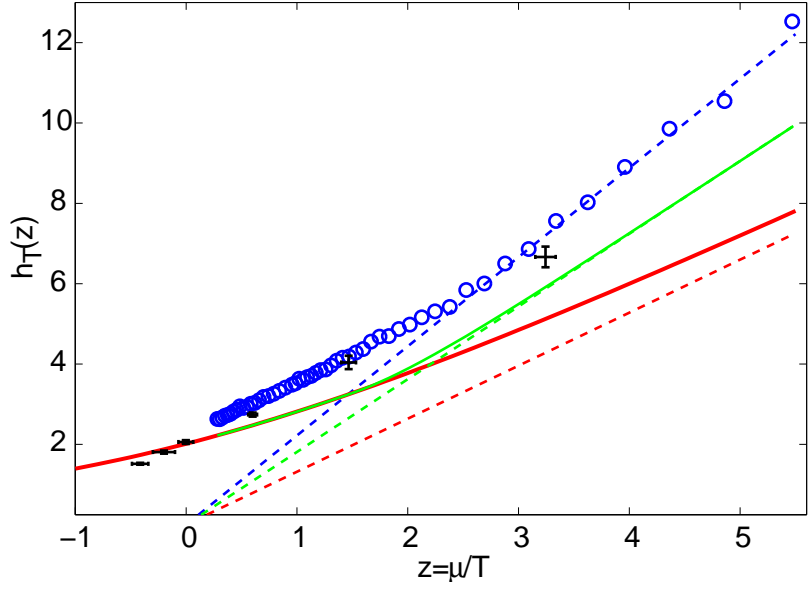}
\caption{\label{hT}(Color online) MC data from (blue circles), Ref.~\cite{burovski} (six black points). The four straight lines starting at the origin are the $T\rightarrow 0$ limits of $h_T(z\rightarrow\infty)=2^{2/5}z/\xi_s^{3/5}$, where $\xi_s=0.42(2)$ \cite{CarlsonReddy, giorgini}, $\xi_s= 0.59$ for meanfield/BCS approximation and $\xi=1$ for the free Fermi gas model respectively. The two solid lines (red/lower and green/higher) correspond to $h_T(z)$ calculated in the free Fermi gas and the BCS/meanfield approximation $h_T(z)$ respectively.}
\end{figure}

We start with the grand-canonical ensemble, where the thermodynamics is derived from the thermodynamic potential $\Omega(T,\mu,V)= - PV$. At unitarity, where both $\mu$ and $T$ are conventionally measured in units of the free gas Fermi energy $\varepsilon_F$, we can write $\Omega$ in terms of a function $h_T(z)$, where $z = \mu/T$, as in Ref.~\cite{TrappedUFG}:
\begin{equation}
\Omega(T,\mu,V) = \Omega(z,V) =-P(z) V = -\frac{2}{5}\beta\left [T h_T(z)\right]^{5/2} V
\end{equation}
where $\beta = \frac{1}{6\pi^2}\left(\frac{2m}{\hbar^2}\right)^{3/2}$.
This form is useful because thermodynamic stability implies three conditions on $h_T$: $h_T > 0$, $h_T' > 0$ and $h_T'' > 0$. The form of this function is shown in Fig.~\ref{hT} along with data of Ref.~\cite{burovski}.

The particle number density and the entropy per particle can then be derived as follows:
\begin{equation}
n = \frac{N}{V} = -\frac{1}{V}\frac{\partial \Omega}{\partial \mu} = \frac{5}{2} \frac{P}{T} \frac{h_T'(z)}{h_T(z)}
\end{equation}
\begin{equation}
\frac{S}{N} = -\frac{1}{N}\frac{\partial \Omega}{\partial T} = \frac{5}{2}\frac{P}{nT}\left[ 1 - z\frac{h_T'(z)}{h_T(z)}\right] = 
\left[\frac{h_T(z)}{h_T'(z)} - z\right]
\end{equation}

Using the thermodynamic identity $E = TS - PV + \mu N$, inserting the expressions above, we find
\begin{equation}
E = \frac{5}{2}PV\left[ 1 - z \frac{h_T'(z)}{h_T(z)}\right]
- PV + z \frac{5}{2} PV \frac{h_T'(z)}{h_T(z)} = \frac{5}{2}PV - PV = \frac{3}{2} PV
\end{equation}
Using this relation together with the constant volume identity $\partial E/\partial T = T\partial S / \partial T$, one can derive relation (\ref{ConsistencyCheck}).

\begin{acknowledgments}

We thank S. Giorgini, A. Gezerlis and J. Carlson, and E. Burovski {\it et al.} for providing us with their data. Support from the Department of Energy under grants DE-FG02-97ER41014 and DE-FG02-00ER41132, from the Polish Ministry of Science under contract No. N N202 328234, and NERSC under grant B-AC02-05CH11231, is gratefully acknowledged. Use of computers at the Interdisciplinary Centre for Mathematical and Computational Modelling (ICM) at Warsaw University is also gratefully acknowledged.
\end{acknowledgments}


\end{document}